\documentclass[twocolumn]{aastex702}

\usepackage{amsmath}
\usepackage{siunitx}
\usepackage{overpic}
\usepackage{comment}
\usepackage{booktabs}

\usepackage{soul}
\newcommand\N[1]{{\color{blue} \bf #1}}

\newcommand{\tabbiblink}[2]{%
  \nocite{#2}%
  \hyperlink{cite.#2}{#1~(\citeyear{#2})}%
  \makebox[0pt][l]{\phantom{\citep{#2}}}%
}

\begin{document}

\title{Quantifying Systematic Age Discrepancies in Very Young Star Clusters}

\author[0000-0001-8878-4994]{Joseph Guzman}
\affiliation{Department of Physics, Florida State University, 77 Chieftan Way, Tallahassee, FL 32306, USA}
\email[show]{jjguzman@fsu.edu}

\author[0000-0003-1599-5656]{Jeremiah W.~Murphy}
\affiliation{Department of Physics, Florida State University, 77 Chieftan Way, Tallahassee, FL 32306, USA}
\email[show]{jwmurphy@fsu.edu}

\author[0000-0003-4666-4606]{Emma Beasor}
\affiliation{Department of Physics, Liverpool John Moores University, 146 Brownlow Hill, Liverpool, UK}
\email[]{E.R.Beasor@ljmu.ac.uk}

\author[0000-0002-1264-2006]{Julianne J.~Dalcanton}
\affiliation{Astronomy Department, University of Washington, Box 351580, Seattle, WA 98195-1580, USA}
\affiliation{Center for Computational Astrophysics, Flatiron Institute, 162 Fifth Avenue, New York, NY 10010, USA}
\email[]{}

\author[0000-0001-5510-2424]{Nathan Smith}
\affiliation{Steward Observatory, University of Arizona, 933 N. Cherry Ave., Tucson, AZ 85721, USA}
\email[]{}

\author[0000-0001-8341-3940]{Mojgan Aghakhanloo}
\affiliation{Department of Astronomy, University of Virginia, 530 McCormick Road, Charlottesville, VA 22904, USA}
\affiliation{Virginia Institute of Theoretical Astronomy, University of Virginia, Charlottesville, VA 22904, USA}
\email[]{mvy4at@virginia.edu}

\author[0000-0002-7502-0597]{Benjamin F.~Williams}
\affiliation{Astronomy Department, University of Washington, Box 351580, Seattle, WA 98195-1580, USA}
\email[]{}

\author[0000-0001-8196-7229]{Andr\'es F.~Barrientos}
\affiliation{Department of Statistics, Florida State University, 117 N. Woodward Ave., Tallahassee, FL 32306, USA}
\email[]{}

\begin{abstract}
We infer the ages of three young stellar clusters—NGC~2004, NGC~7419 and NGC~2100—using Stellar Ages, a statistical algorithm designed to infer stellar population properties from color–magnitude diagrams. Recent studies have revealed emerging inconsistencies in the inferred ages of very young ($\le$ 50 Myr-old) stellar clusters. Here, we identify and quantify two distinct discrepancies. First, ages inferred from red supergiants (RSGs) are systematically older—by about a factor of four—than those inferred from bright blue stars, with a median age offset of $\Delta \log_{10}(t/{\rm yr}) = 0.60 \pm 0.06~{\rm dex}$ ($t_B \approx t_{\rm RSG}/4.0$). Second, given the observed numbers of RSGs and bright blue stars, there is a pronounced deficit of lower-mass main-sequence stars relative to expectations from a standard initial mass function. Although these discrepancies resemble those reported for intermediate-age clusters, their magnitude and character suggest that they are unique to the evolution of massive stars. Together, these results highlight population-level inconsistencies with single-star evolutionary models and underscore the need to consider multiple evolutionary tracers when age-dating young clusters. By combining individual stellar ages with population-wide constraints, our approach refines prior work on cluster age determinations and provides new insight into massive star evolution and the interpretation of cluster demographics.
\end{abstract}


\section{Introduction}
\label{sec:intro}

Star clusters have long served as testing grounds for stellar evolution theory, under the assumption that they form as simple stellar populations (SSPs), groups of stars with a common age and metallicity \citep{Tinsley1972, Bruzal2003}. This assumption makes them valuable benchmarks for testing stellar models. However, past work has often shown that SSP assumptions can break down \citep{Bastian_2006, Piotto2007}. Color–magnitude diagrams (CMDs) of young and intermediate-age clusters often reveal features that are difficult to reconcile with a single-age population, including multiple main sequences, and stars more luminous than the nominal turnoff \citep{Mackey2007, Li_2017}. These features complicate cluster age measurements while highlighting limitations in stellar evolution models.

One example is the study of \citet{Beasor2019}, who analyzed four young clusters: NGC 2004 and NGC 2100 in the Large Magellanic Cloud (LMC), and NGC 7419 and $\chi$ Persei in the Milky Way. By fitting isochrones to different CMD features, they found a wide range of inferred ages, from about 7 to 23 Myr. The choice of stellar tracer strongly affected the results: lowest luminosity RSGs, main-sequence turnoff stars, and luminosity functions did not yield consistent ages. They suggested that binary interactions or mergers could produce apparently rejuvenated stars, mimicking a younger population \citep{deMink2014, Schneider2014, Britavskiy2019}.

This tension is not unique to the Magellanic clusters. A wide body of work has shown that apparent age spreads in young clusters may reflect physical processes other than prolonged star formation. Binary interactions are now known to be common among massive stars, with multiplicity fractions exceeding 70\% \citep{HSana2012, Moe2017}. These interactions can create rejuvenated mass gainers or merger products that appear as anomalously young bright blue stars, qualitatively analogous to blue stragglers in older populations \citep{Sandage1953, Ferraro2009}. Rotation offers an additional channel for extending main-sequence lifetimes and altering luminosities \citep{Maeder2000, Ekstrom2012_Geneva}. Rapid rotators, such as Be stars, are particularly prevalent in young clusters and may contribute to the observed CMD complexities \citep{porter2003}. These mechanisms not only complicate cluster age-dating, but also bear directly on massive star evolution. Understanding how cluster demographics are shaped by binaries and rotation therefore informs both stellar evolution modeling and confidence in mass estimates for core-collapse supernova progenitors.

A variety of statistical methods have been developed for CMD analysis, ranging from isochrone fitting of individual stars to synthetic CMD approaches that reconstruct star-formation histories \citep[e.g.,][]{Harris_2001_StarFISH, dolphin2002, daSilva2006_PARAM, Perren2015_Asteca}. These frameworks have proven effective in many contexts, yet they often face challenges in young clusters where only a small number of evolved stars dominate the age signal. In such regimes, traditional CMD binning approaches can yield unstable results, while methods tailored to individual stars often neglect the population context. The Stellar Ages algorithm builds on this body of work by combining individual-star age estimates with population-wide inference, enabling consistency tests across distinct stellar tracers.

In this paper, we revisit three of these clusters with a new statistical approach. We apply Stellar Ages \citep{Guzman_2025}, a Bayesian algorithm designed to infer population properties directly from CMDs. Unlike traditional isochrone overlaying techniques, Stellar Ages provides posterior age distributions for both individual stars and the ensemble, enabling systematic comparisons between different stellar tracers. By testing whether populations such as RSGs, main-sequence stars, and bright blue stars yield consistent age distributions, we identify systematic discrepancies that reveal where current single-star models fail to reproduce the observed diversity in very young clusters.

The paper is organized as follows: in Section~\ref{sec:data} we describe the photometric datasets, sample selection, and adopted cluster parameters; Section~\ref{sec:model} outlines the Stellar Ages algorithm; Section~\ref{sec:results} presents the results, including inferred age distributions, population mismatches, and the role of bright blue stars; Section~\ref{sec:Interpretation} interprets these findings in the context of previous work, stellar evolution models, and binary evolution; and Section~\ref{sec:conclusions} summarizes our conclusions.

\section{Data and Clusters}
\label{sec:data}

The goal of this study is to compare age estimates for distinct stellar age tracers, RSGs, main sequence (MS) stars, and bright blue stars, in three young nearby clusters. The sample consists of two clusters in the LMC (NGC 2100 and NGC 2004) and one in the Milky Way (NGC 7419). \citet{Beasor2019} also analyzed $\chi$ Persei using the Kitt Peak photometry of \citet{Currie2010_ChiPersei}, but the Kitt Peak dataset differs substantially from the others in both filter coverage and depth, and does not provide a complete or homogeneous sample of massive stars. After quality cuts, the photometry of $\chi$ Persei contained too few bright stars for a stable Stellar Ages analysis, which undercuts a meaningful comparison across our age tracers. For these reasons we did not include $\chi$ Persei in our final sample.

To ensure direct comparability, we rely, where possible, on the same archival photometric datasets used in \citet{Beasor2019}.

\subsection{Archival Photometry}
\label{subsec: MS Phot}
For the LMC clusters NGC~2100 and NGC~2004, we adopt the $B$- and $V$- band de-reddened photometry from~\citet{Niederhofer2015}, originally compiled from~\citet{Brocato2001}. The Niederhofer et al. study is based on archival imaging with the Hubble Space Telescope Wide Field and Planetary Camera 2 (WFPC2). The HST WFPC2 observations underlying the photometry for NGC~2100 and NGC~2004 were obtained under HST Program~5475 (PI: Shara) from the Mikulski Archive for Space Telescopes (MAST) at the Space Telescope Science Institute. The specific observations analyzed can be accessed via~\dataset[https://doi.org/10.17909/s913-gy65]{https://doi.org/10.17909/s913-gy65}. For NGC~7419, we adopt $UBV$ photometry from~\citet{Beauchamp1994_NGC7419}, acquired using direct CCD imaging at the Mount M\'egantic Observatory between 1987 and 1989.

A key requirement of our Stellar Ages analysis is that the photometric catalog samples the main sequence and evolved populations in a self-consistent manner. Incomplete sampling or spatially mismatched samples can bias inferred cluster properties, because the algorithm is sensitive to the relative numbers of stars in different evolutionary phases. To mitigate these effects, we follow the completeness and spatial selection criteria established in the original photometric studies.~\citet{Brocato2001} performed artificial-star tests and demonstrated that the photometry is at least 90 percent complete down to $V\approx19.6$ for NGC~2100 and $V\approx19.8$ for NGC~2004 within three times the cluster core radius.\citet{Niederhofer2015} further restricted their analysis to stars located within twice the core radius of each cluster. We adopt this same two-core radius spatial criterion and apply additional quality cuts, retaining only stars with photometric uncertainties of 0.1 mag or less. Together, these selections ensure that the resulting main-sequence catalogs are both spatially well defined and photometrically reliable.

For NGC~7419, we adopt the $UBV$ photometry of~\citet{Beauchamp1994_NGC7419}, which was obtained from direct CCD imaging at the Mount M\'egantic Observatory. Subsequent studies of this cluster have emphasized strong and spatially variable extinction across the field, while preserving the census of bright evolved stars originally identified in the Beauchamp et al. catalog. \citet{Joshi_2008} and \citet{Marco_2013} reaffirmed the presence of the five RSGs, and this RSG population has recently been confirmed in Gaia-based membership studies \citep[e.g.,][]{Chakraborty_2025}. These later works did not perform formal completeness tests. As for the LMC clusters, we retain only stars with photometric uncertainties of 0.1 mag or less. Accordingly, to ensure a repeatable comparison with \citet{Beasor2019}, we adopt the~\citet{Beauchamp1994_NGC7419} dataset as the basis for our analysis of NGC~7419.

\begin{figure*}[htbp]
    \centering
    \includegraphics[width=0.45\textwidth]{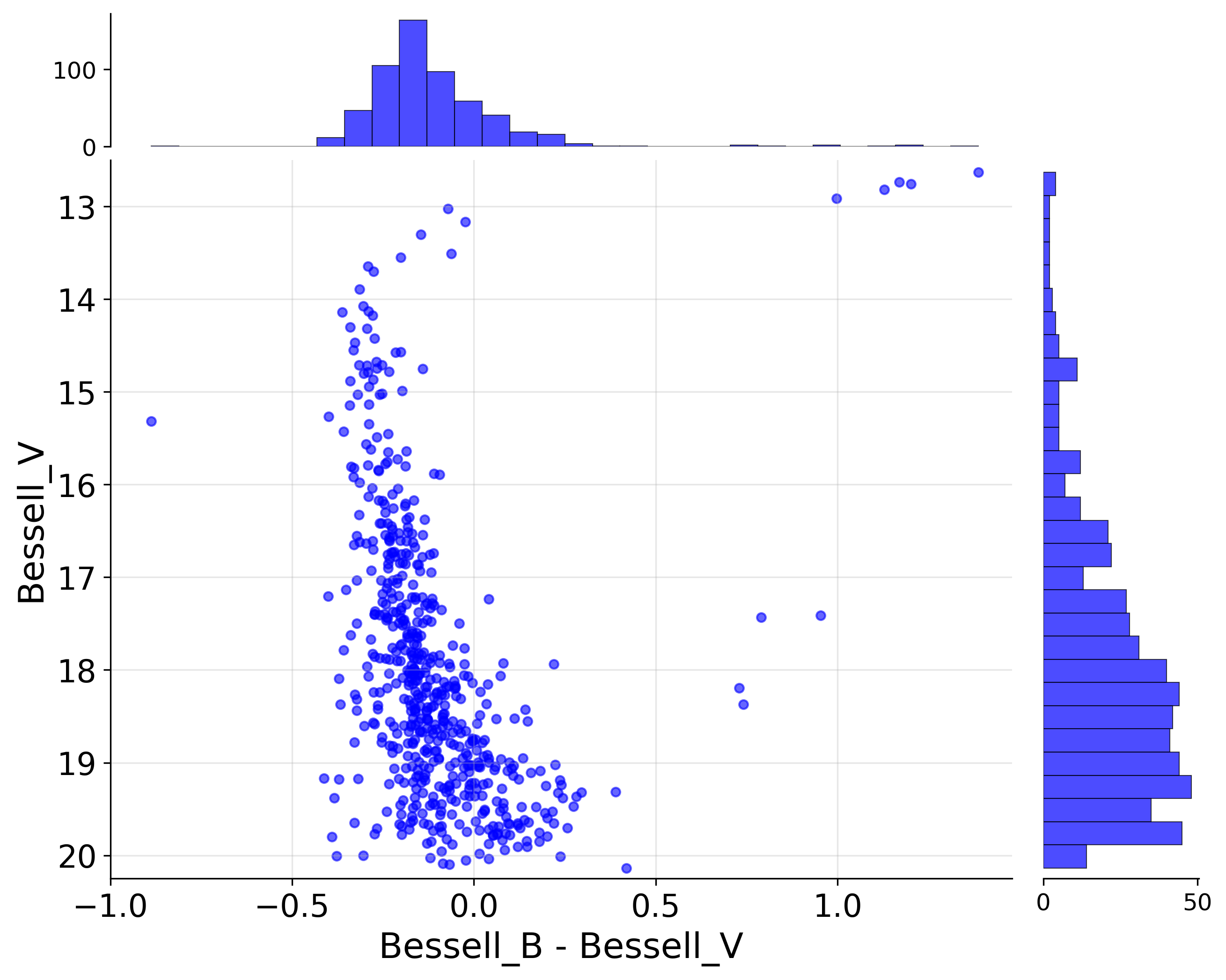}\hfill
    \includegraphics[width=0.45\textwidth]{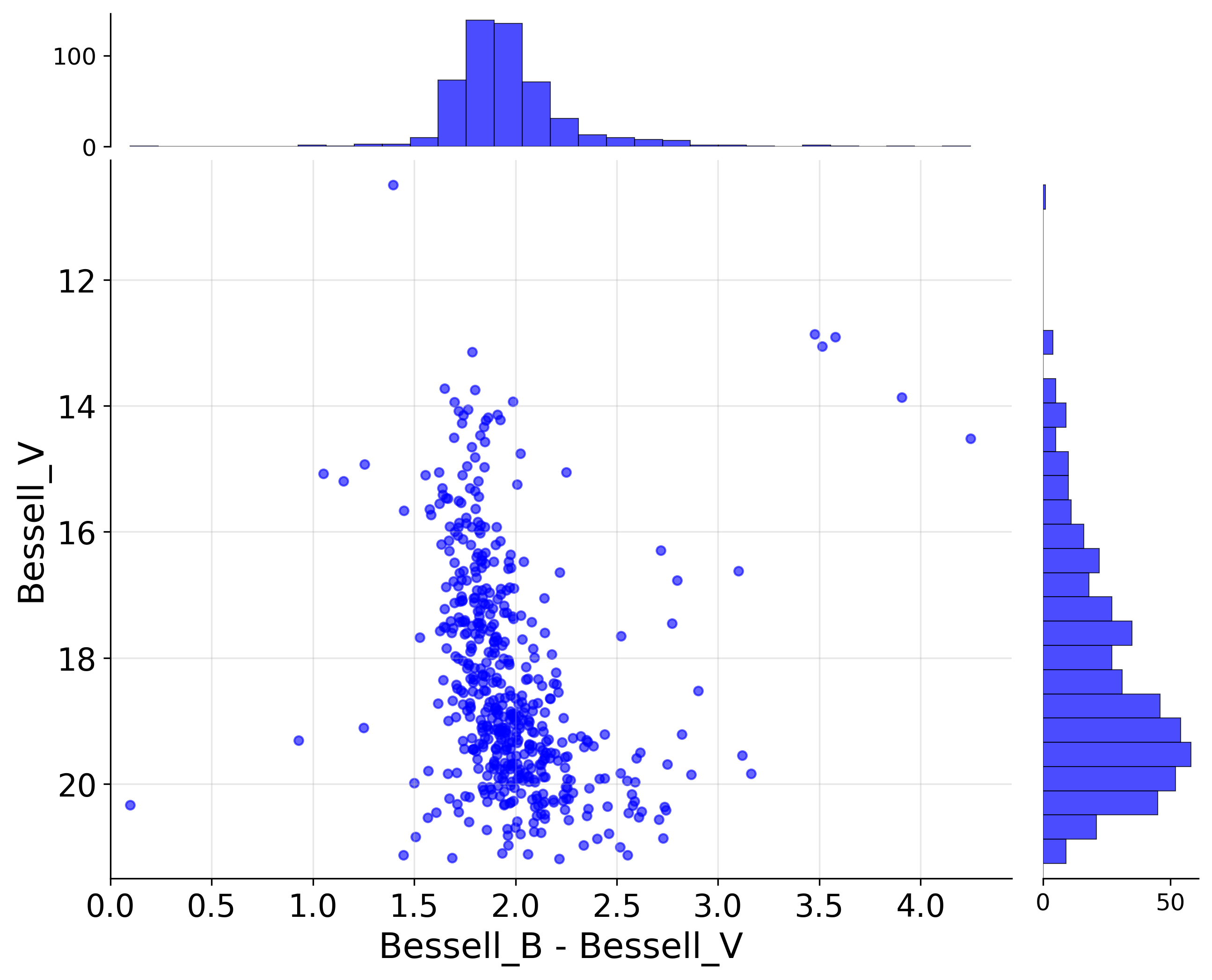}\\[6pt]
    \includegraphics[width=0.45\textwidth]{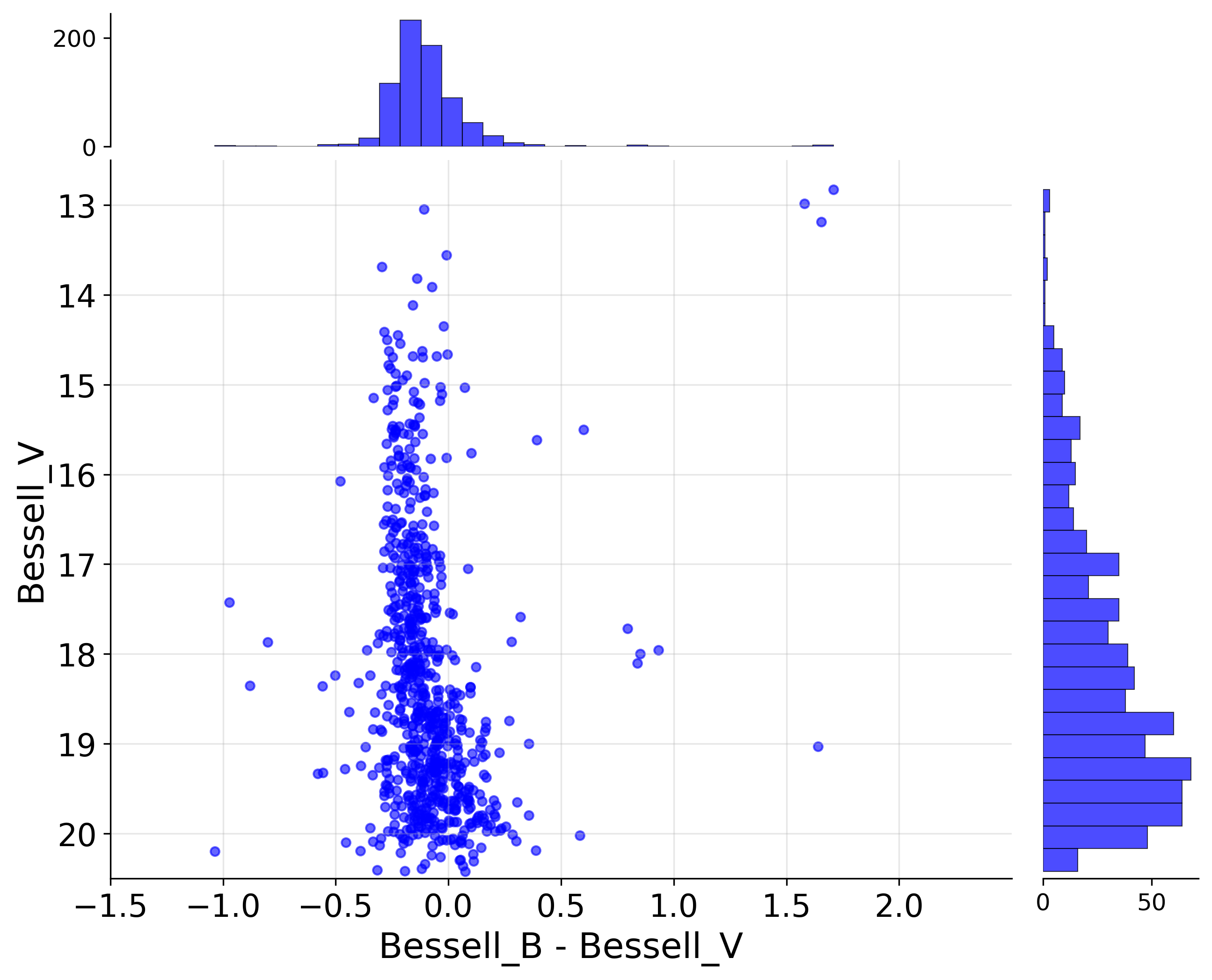}
    \caption{\small CMDs of cluster stars in NGC~2004 (top left), NGC~7419 (top right), and NGC~2100 (bottom). The LMC clusters NGC~2004 and NGC~2100 include de-reddening corrections following \citet{Niederhofer2015}, whereas NGC~7419 is shown without extinction correction.  Here, we present the full data with no magnitude limits. In subsequent figures and analysis, we apply magnitude limits to retain stars with magnitude uncertainty of 0.1 or less across all three clusters. The top panels present a histogram across $B\!-\!V$, while the right panels present a histogram across $V$ magnitudes.}
    \label{fig:3CMDs}
\end{figure*}

For reference, we present the full color–magnitude diagrams of NGC~2004, NGC~7419, and NGC~2100 in Figure \ref{fig:3CMDs}. These diagrams show the complete photometric samples prior to the application of magnitude-uncertainty cuts used in the analysis below.

The presence of strong and spatially variable extinction in NGC~7419 motivates us to model reddening more broadly than a single fixed value, a point to which we return in the following two sections.

\subsection{Distances and Extinction}
\label{subsec: Distances and Extinction}
For the LMC clusters, we adopt a fixed distance of 50 kpc from \citet{Pietrzynski2013_LMCDistance}. For NGC~7419, we adopt a distance of 2.93 kpc, as derived in \citet{Davies2019}.

\begin{figure}[ht]
    \centering
    \includegraphics[width=\columnwidth]{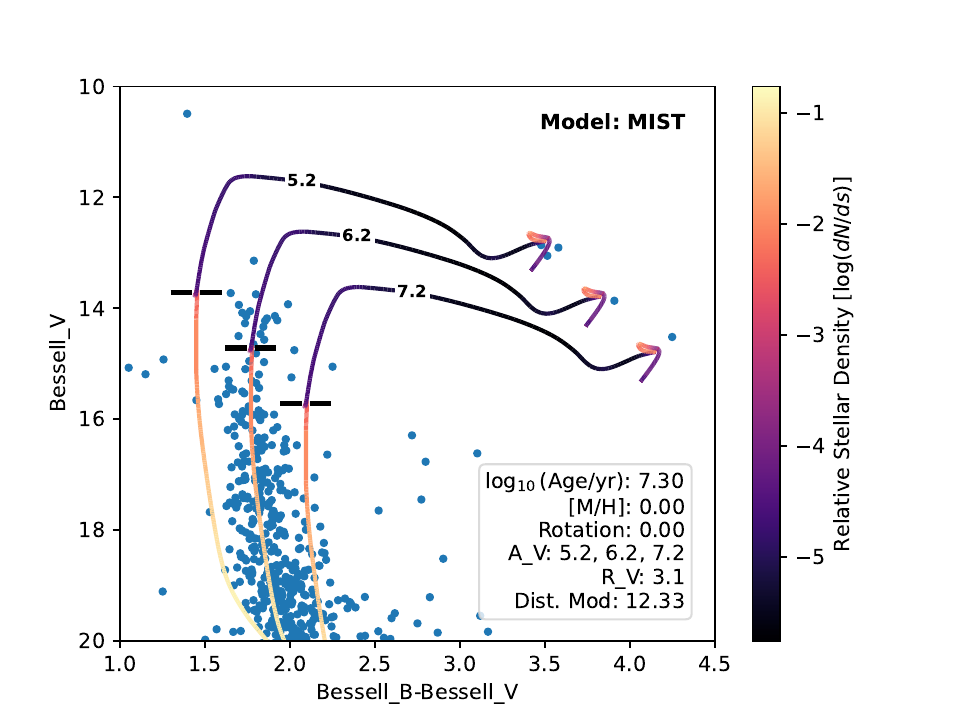}
    \caption{\small CMD for cluster NGC~7419 and MIST isochrones showing three different extinctions.  The color map shows the relative density of stars assuming a Salpeter IMF.  The horizontal marks indicate the MSTO defined by the drop in stellar density (where $\log_{10}(dN/ds)$ drops below -3). This preliminary analysis strongly indicates that the median extinction is around $A_V = 6.2$ but that there is also a wide distribution of extinctions, which we later model with a log-normal distribution.}
    \label{fig:CMDAvDemo}
\end{figure}

Extinction treatment follows the published analyses of each cluster. For NGC~2004,\citet{Niederhofer2015} applied a uniform reddening correction of $E(B-V)=0.23$, while for NGC~2100 they explicitly corrected the photometry for differential extinction across the cluster field using the method of \citet{Milone2012_reddening}. For NGC~7419, \citet{Beasor2019} reported a mean extinction of $\langle A_{V} \rangle = 6.33$, which is somewhat consistent with the extinction estimate provided by \citet{Beauchamp1994_NGC7419} of $A_{V} = 6.2$.  On the other hand, \citet{Marco_2013} report a mean extinction of $A_V = 5.2$. To assess this discrepancy, Figure~\ref{fig:CMDAvDemo} presents a preliminary analysis, showing the effects of extinction on a few representative isochrones. The age, metallicity, and rotation are representative of parameters that are consistent with RSG magnitudes.The points show the CMD for NGC~7419 and each of the three curves shows an isochrone with a different extinction: $A_{V} = 5.2$, 6.2, and 7.2.  This figure illustrates two facts: first, given the choice of this specific age, metallicity, and rotation, the median extinction is closest to 6.2 not 5.2; second, there is clearly a wide distribution of extinctions. These isochrones are for illustration purposes; later we will infer the most likely age using a full range of models and a log-normal distribution for $A_V$. The details of this implementation are described in section \ref{sec:model}.

\subsection{Published Age Estimates for the Target Clusters}
\label{subsec:literature_ages}

Before describing our methodology, it is useful to summarize the range of published age estimates for the three clusters analyzed in this work. Table~\ref{tab:cluster_ages_lit} compiles representative age determinations for NGC~2004, NGC~2100, and NGC~7419 drawn from the literature. These studies employ a variety of techniques, including main-sequence turn-off fitting, integrated photometry, ultraviolet spectroscopy, and red supergiant based methods.

\begin{deluxetable*}{lccc p{4.2cm} l}
\tabletypesize{\scriptsize}
\tablecaption{Literature Age Estimates for NGC~2004, NGC~7419, and NGC~2100} \label{tab:cluster_ages_lit}
\tablehead{
\colhead{Reference} &
\colhead{NGC~2004} &
\colhead{NGC~7419} &
\colhead{NGC~2100} &
\colhead{Method}
}
\setlength{\tabcolsep}{4pt}
\startdata
\citet{Bencivenni1991}        & 8                 & \nodata              & \nodata           & CMD fitting (UBV) \\
\citet{Elson1991}                     & 20                & \nodata              & 16                & Integrated photometry and CMD \\
\tabbiblink{L. A. Balona}{Balona1993_1}
    & 40                & \nodata              & 40            & CMD fitting (UBV) \\
\citet{Bhatt1993}             & \nodata           & 40                   & \nodata           & CMD fitting (UBV) \\
\citet{Beauchamp1994_NGC7419}         & \nodata           & $14 \pm 2$           & \nodata           & CMD fitting (UBV) \\
\citet{Caloi1995}       & 20                & \nodata              & \nodata           & UV spectroscopy (IUE) \\
\citet{Cassatella1996}        & \nodata           & \nodata              & 15                & UV spectroscopy (IUE) \\
\citet{Subramaniam2006}       & \nodata           & $25 \pm 5$           & \nodata           & MS turn-off isochrones \\
\citet{Joshi_2008}             & \nodata           & $22.5 \pm 3^{a}$         & \nodata           & Turn-off and morphological age \\
\tabbiblink{A. Marco et al.}{Marco_2013}       & \nodata           & $14 \pm 2$           & \nodata           & Str\"omgren photometry + spectra \\
\tabbiblink{F. Niederhofer et al.}{Niederhofer2015}       & $20 \pm 1.4$      & \nodata              & $19.5 \pm 2$                & HST CMD and isochrone fitting \\
\citet{Patrick2016}           & \nodata           & \nodata              & $20 \pm 5$        & Geneva isochrones + RSGs \\
\citet{Beasor2016_NGC2100RSGs}          & \nodata           & \nodata              & 14                & RSG luminosity function \\
\citet{Beasor2019}            & 7--24             & 6--24                & 7--21             & MSTO and RSG isochrones \\
\citet{Chakraborty_2025}       & \nodata           & $21.1^{+1.6}_{-0.6}$ & \nodata           & Gaia CMD and astrometry \\
\enddata
\tablecomments{Ages are reported in Myr and listed as given in the original references. MSTO = main-sequence turn-off; IUE = \textit{International Ultraviolet Explorer}. $^{a}$ \citet{Joshi_2008} also report a morphological age range of 20--25 Myr.}
\end{deluxetable*}

Although all three clusters are consistently classified as young systems with ages of order tens of Myr, the reported values span a wide range. In several cases, different tracers applied to the same cluster yield discrepant ages that differ by factors of two or more. This spread is most evident when comparing ages inferred from the main-sequence turn-off to those derived from red supergiant populations, as highlighted in recent studies of young clusters.

The diversity of methods and resulting age estimates underscores two points that motivate the present analysis. First, cluster age determinations remain sensitive to the specific stellar population used as the age tracer. Second, it is not always clear which age estimate best reflects the underlying stellar population when different tracers disagree. These ambiguities motivate a unified statistical framework that can compare multiple stellar populations on equal footing, which we address using the Stellar Ages algorithm described in the following section.

\section{Method: The Stellar Ages Algorithm}
\label{sec:model}

To determine the age distributions of NGC 2004, NGC 2100, and NGC 7419, we apply Stellar Ages \citep{Guzman_2025}, a probabilistic framework designed to infer the underlying distributions of age, metallicity, rotation, and extinction directly from individual stellar photometry. Unlike most population-fitting methods, this approach provides age likelihoods not only for the ensemble, but also for each star in the sample.

Traditional color–magnitude diagram (CMD) fitting methods reconstruct a star formation history by dividing the CMD into bins and comparing the observed star counts in each bin to those from synthetic populations (e.g., \citet{dolphin2002, dolphin2013, Gallart2005_SynthCMD, Aparicio_2009_SynthCMD}). The bin counts are typically modeled as independent Poisson draws, and the likelihood is the product over all the bins. While powerful for large stellar systems, this formulation is vulnerable to small-number statistics in the regimes relevant to massive star and supernova progenitor studies. In such clusters, the small handful of luminous, evolved stars (RSGs, blue supergiants, or luminous blue variables) carry most of the age information. Yet in a Poisson-bin framework, the likelihood can hinge on whether one or two stars fall in a bin, rather than by the astrophysically significant fact that such luminous stars exist. A more precise strategy is to reframe the statistical question from `How many stars do we expect in a bin for this age?' to `What is the most likely age of each individual star, given the context of the full population?'

Stellar Ages adopts this strategy by modeling the full joint probability distribution of stellar magnitudes without binning. For each star, the method evaluates the likelihood of its observed photometry across a precomputed grid of model parameters (age, metallicity, rotation, extinction), to construct an age likelihood function tailored to that star. These likelihoods are then combined in a hierarchical Bayesian framework, allowing the brightest, most age-sensitive stars to sharpen the inference while still incorporating the statistical weight of the main-sequence majority. In this way, the approach merges the strengths of isochrone fitting and synthetic CMD methods, yielding robust constraints in clusters where only a few evolved stars dominate the age signal.

This design has two key advantages. First, it preserves the strong age constraints offered by the most luminous stars. For example, a single star with $M_{\mathrm{F814W}}=-8.6$ almost certainly implies an age under $\sim$10 Myr and an initial mass above $\sim$19 M$_\odot$, so only a few such stars are required to provide tight constraints on the cluster age. Second, it reframes the statistical problem away from expected star counts toward the most likely age of each star in the context of the population, thereby reducing the impact of small-number statistics. By providing age likelihoods for both individual stars and the ensemble, Stellar Ages allows us to check the internal self-consistency across the CMD.

In general, Stellar Ages infers the posterior distribution for age ($t$), metallicity ([M/H]), initial rotation as a fraction of the critical rotation ($v_{\rm ini}$), and the median extinction, $\tilde{A}_{V}$. However, since we will treat $A_{V}$ as a prior from \citet{Beasor2019}, we do not need to infer $\tilde{A_{V}}$. Therefore, the model parameters are $\theta = \{t, \rm{[M/H]}, v_{\rm{ini}} \}$. Schematically, the posterior distribution is as follows:
\vspace{-0.1cm}
\begin{equation}
P(\theta | D) \propto \mathcal{L}(D | \theta) P(t) P(\rm{[M/H]}) P(v_{\rm{ini}}) \, .
\end{equation}

The data ($D$) are the set of magnitudes for all stars.  The modeled magnitude in band ($a$) of each star is
\begin{equation}
    m_{a} = \tilde{m}_{a}(M,t,\rm{[M/H]},v_{\rm{ini}}) + A_a + e_{a} \, ,
    \label{eq:maga}
\end{equation}
where $\tilde{m}_a$ is the magnitude predicted by stellar evolution models as a function of $t$, [M/H], and the initial mass ($M$) of the star \citep[MIST]{MIST2016_0, MIST2016_1}.  $e_a$ is a random error, drawn from a Gaussian distribution with width $\sigma_a$; $\sigma_a$ is the magnitude uncertainty. $A_{a}$ represents the extinction parameter in band $a$. In this analysis, we model two magnitudes, which are the Bessell apparent B, and V magnitudes.

In this work, we adopt the MIST isochrones \citep{MIST2016_0, MIST2016_1}, which provide predicted B and V magnitudes over a grid of age, metallicity, and initial rotation. This choice maintains consistency with \citet{Beasor2019} and offers coverage up to $\sim$300 M$_\odot$ and rotation fractions of $v{\rm ini}/v_{\rm crit}=0.0$ and $0.4$, spanning the full mass range relevant to this study. Although we use MIST here, the Stellar Ages framework is compatible with any stellar evolutionary grid that supplies synthetic magnitudes. For instance, \citet{Guzman_2025} employed PARSEC v1.2S \citep{Chen2015, Marigo_2017, Fu2018}, while \citet{Murphy_2025} used both PARSEC v2.0 \citep{Costa2019a_Parsec2.0, Costa2019b_Parsec2.0, Nguyen2022_Parsec2.0} and MIST. A limitation of PARSEC v2.0 and Geneva models \citep{Lejeune2001_Geneva, Ekstrom2012_Geneva, Eggenberger2021_GenevaRot} is that rotation coverage is restricted to a narrower range of masses and ages, complicating inference for young, massive populations. In \citet{Murphy_2025}, we showed that the differences between MIST and PARSEC v2.0 results were smaller than the statistical variation introduced by a 0.1 dex shift in $\log_{10}$(t/yr). For these reasons, we base our analysis on MIST isochrones with rotation, while noting that Stellar Ages can readily incorporate alternative models as their parameter coverage improves.

The joint likelihood of observing a star with magnitudes $m_a$ and $m_b$ is 
\vspace{-0.3cm}
\begin{multline}
 \mathcal{L}(m_a,m_b|\theta) = \int p(m_{a}|\tilde{m}_{a})p(m_{b}|\tilde{m}_{b})p(\tilde{m}_{a}|\theta,M) \\
 p(\tilde{m}_{b}|\theta,M) p(M) dM d\tilde{m}_{a} d\tilde{m}_{b} \, . 
\end{multline}
This likelihood is the joint probability density function for a given age and metallicity.  $p(m_{a}|\tilde{m}_{a})$ and $p(m_{b}|\tilde{m}_{b})$ are Gaussian distributions whose widths represent the observational uncertainties.  $p(\tilde{m}_a | \theta, M)$ and $p(\tilde{m}_b | \theta, M)$ are delta functions and come directly from stellar evolution.  Since $\tilde{m}_a$ and $\tilde{m}_b$ are not analytic, there is no closed analytic form for the joint probability density function.  To approximate the likelihood, we note that the likelihood is the expectation of $p(m_a | \tilde{m}_a)$ and $p(m_b | \tilde{m}_b)$ with respect to the initial mass $M$.  Therefore,

\begin{multline}
    \mathcal{L}(m_a,m_b|\theta) = 
    E_M \left [ p(m_a | \tilde{m}_a(\theta, M)) 
    p(m_b | \tilde{m}_b(\theta, M)) \right ] \\
    \approx \frac{1}{N_M} \sum^{N_M}_\ell 
    p(m_A | \tilde{m}_A(\theta, M^{(\ell)}))
    p(m_B | \tilde{m}_B(\theta, M^{(\ell)})) \, ,
\end{multline}
where each $M^{(\ell)}$ is a draw from $P(M)$, the initial mass function.  To minimize sampling noise, we use perfect sampling instead of random sampling when drawing from this distribution.

Real stellar populations will likely be a mixture of populations with different ages and metallicities. Therefore, we propose a mixture data-generating model, or a weighted sum of the individual joint probability density functions:
\vspace{-0.2cm}
\begin{multline}
\label{mix_model}
    \mathcal{L}(m_A,m_B|\{w_{t,z,v}\}^{T,Z,V}_{t=1,z=1,v=1}) \\
    = \sum^T_t \sum^Z_z \sum^V_v w_{t,z,v} \mathcal{L}(m_A,m_B|\theta_{t,z,v}) \, ,
\end{multline}
where $w_{t,z,v}$ are weights for age index $t$, metallicity index $z$, and initial rotation index $v$.

To infer the weights, we employ a Bayesian framework and use a Gibbs sampler with latent variables ($r_i$). These labels assign specific age, metallicity, and rotation values to each star, making the mixture model (\ref{mix_model}) equivalently
\vspace{-0.1cm}
\begin{align}
\begin{split}
(m_{A,i},m_{B,i} \mid r_i = (t', z', v')) 
&\sim \mathcal{L}(m_A,m_B \mid \theta_{t',z',v'}) \\
r_i = (t', z', v') \mid \{\omega_{t',z',v'}\}
&\sim \omega_{t',z',v'} \quad i = 1,\dots, N_{\rm stars}
\end{split}
\label{mix_model_ri}
\end{align}
These individual labels for each star is what enables an estimate for the age of each star.

We complete the model specification by assigning a prior distribution to the weights. Specifically, we assume that  
\begin{equation}\label{prior_weights}
    \{ w_{t,z,v} \}^{T,Z,V}_{t=1,z=1,v=1} \sim {\rm Dirichlet}(1, ...,1) \, 
\end{equation}
where $\rm{Dirichlet}(1, ...,1)$ denotes a $(T \times Z \times V)$-dimensional Dirichlet distribution. Under model (\ref{mix_model_ri}) with prior (\ref{prior_weights}), the Gibbs sampler iterates between updating the labels and weights.

To update the labels, we sample from the conditional distribution 
\vspace{-0.2cm}
\begin{multline}
P(r_i = (t',z',v')| \{ w_{t,z,v} \}^{T,Z,V}_{t=1,z=1,v=1}) \\ = \frac{w_{t',z',v'} p(m_{a,i},m_{b,i} | \theta_{t',z',v'}) }{ \sum_{t} \sum_{z} \sum_v w_{t,z,v} p(m_{a,i},m_{b,i}|\theta_{t,z,v})} \,  \quad (i = 1...N_{\rm stars}) \, .
\end{multline}

To update the weights, we first compute the number of $r_i$, $i = 1...N_{\rm stars}$, that are equal to $(t',z',v')$ denoting the resulting count as $N(t=t',z=z',v=v')$. Then, we update the weights by sampling from the Dirichlet distribution:

\vspace{-0.6cm}

\setlength{\belowdisplayskip}{2pt}
\begin{multline}
    \{ w_{t,z,v} \}^{T,Z,V}_{t=1,z=1,v=1} | \{r_i\}_{i=1}^{N_{\rm stars}} \\
    \sim {\rm Dirichlet}[1+N(t=1,z=1,v=1) \\,...,1+N(t=T,z=Z,v=V)] \,
\end{multline}
\setlength{\belowdisplayskip}{12pt}

For NGC 7419, we incorporate the external extinction constraint of $\langle A_{V} \rangle = 6.33$ from \citet{Beasor2019} by modeling the distribution as log-normal, following \citet{Dalcanton_2015}. In this parameterization, $\tilde{A}_{\text{V}}$ represents the median extinction and $\sigma_{\text{V}}$ sets the width and skewness of the distribution.

The log-normal distribution is given by:

\vspace{-0.6cm}
\setlength{\belowdisplayskip}{0pt}
\begin{align}
p(A_{\text{V}}|\tilde{A}_{\text{V}}, \sigma_{\text{V}})dA_{\text{V}} &= \frac{1}{A_{\text{V}}\sqrt{2\pi \sigma^{2}_{\text{V}}}}\nonumber\\
&\quad \times \exp\left(\frac{-(ln(A_{\text{V}}/\tilde{A}_{\text{V}}))^{2}}{2\sigma^{2}_{\text{V}}}\right)dA_{\text{V}}
\label{eqn:Av}
\end{align}
\setlength{\belowdisplayskip}{12pt}

We adopt $\sigma_{\text{V}} = 0.3$ based on \citet{Kainulainen2009_MWExtinction}'s Milky Way measurements, and verified consistent results using $\sigma_{\text{V}} = 0.24$ from \citet{Dalcanton_2015}'s M31 analysis. 
To convert between the mean extinction $\langle A_{V} \rangle$ measurement and the sampled median extinction $\tilde{A_{V}}$, we apply eq.~(6) from \citet{Dalcanton_2015}, i.e.

\setlength{\belowdisplayskip}{0pt}
\begin{equation}
    \langle A_{V} \rangle = \tilde{A_{V}} \, e^{\sigma_{\scriptscriptstyle V}^{2}/2}
\end{equation}
\setlength{\belowdisplayskip}{12pt}

If you would like to consider the actual width of the log-normal in magnitudes of extinction, see eq.~(7) in \citet{Dalcanton_2015}.
For more discussion on the techniques and validation tests of Stellar Ages, see \citet{Guzman_2025}.

\section{Results}
\label{sec:results}

In the following section, we present the general inference results, and we further elucidate the inconsistencies of the results.  Section~\ref{subsec:CMD Weights} presents the general results, which include the posterior distributions of age, metallicity, and rotation across each cluster; this section also identifies the primary inconsistency between the ages inferred from the RSGs and bright blue stars. In Section~\ref{subsec:RSGs}, we further quantify this discrepancy by comparing the main-sequence populations predicted by each evolved-star tracer with the observed stellar counts. To illustrate that the bright blue stars and bright red stars are photometrically inconsistent, we then infer ages independently from RSG and bright-blue-star subsets in Section~\ref{subsec: siblings}.  The inferences for luminosity and temperature imply spectral types and luminosity classes; we therefore compare the photometrically derived properties with available spectroscopic constraints in Section~\ref{subsec:spectra}. Finally, Section~\ref{subsec:stat sum} summarizes the resulting differences in cluster age and implied maximum stellar mass across the three clusters.

\subsection{Posterior Age Distributions Across the CMD and an Emerging Inconsistency}
\label{subsec:CMD Weights}

To infer the posterior age distributions for each cluster, we used the Stellar Ages pipeline in \texttt{tzw} mode (age, metallicity, rotation). In the following analysis, the age parameter grid spans ages from $\log_{10}$(t/yr) = 6.50 to 9.00 (3.2 to 1000 Myr), the metallicities span [M/H] = -0.60 to 0.00, and initial rotation rates include two values: 0.0 and 0.4 (constrained by available MIST isochrones). The MCMC sampling contains 5000 steps with a burn-in period of 50 steps, followed by thinning every 10th sample to mitigate autocorrelation effects and ensure statistical independence of the posterior samples.

\begin{figure*}[htbp]
    \centering
    \includegraphics[width=0.325\textwidth]{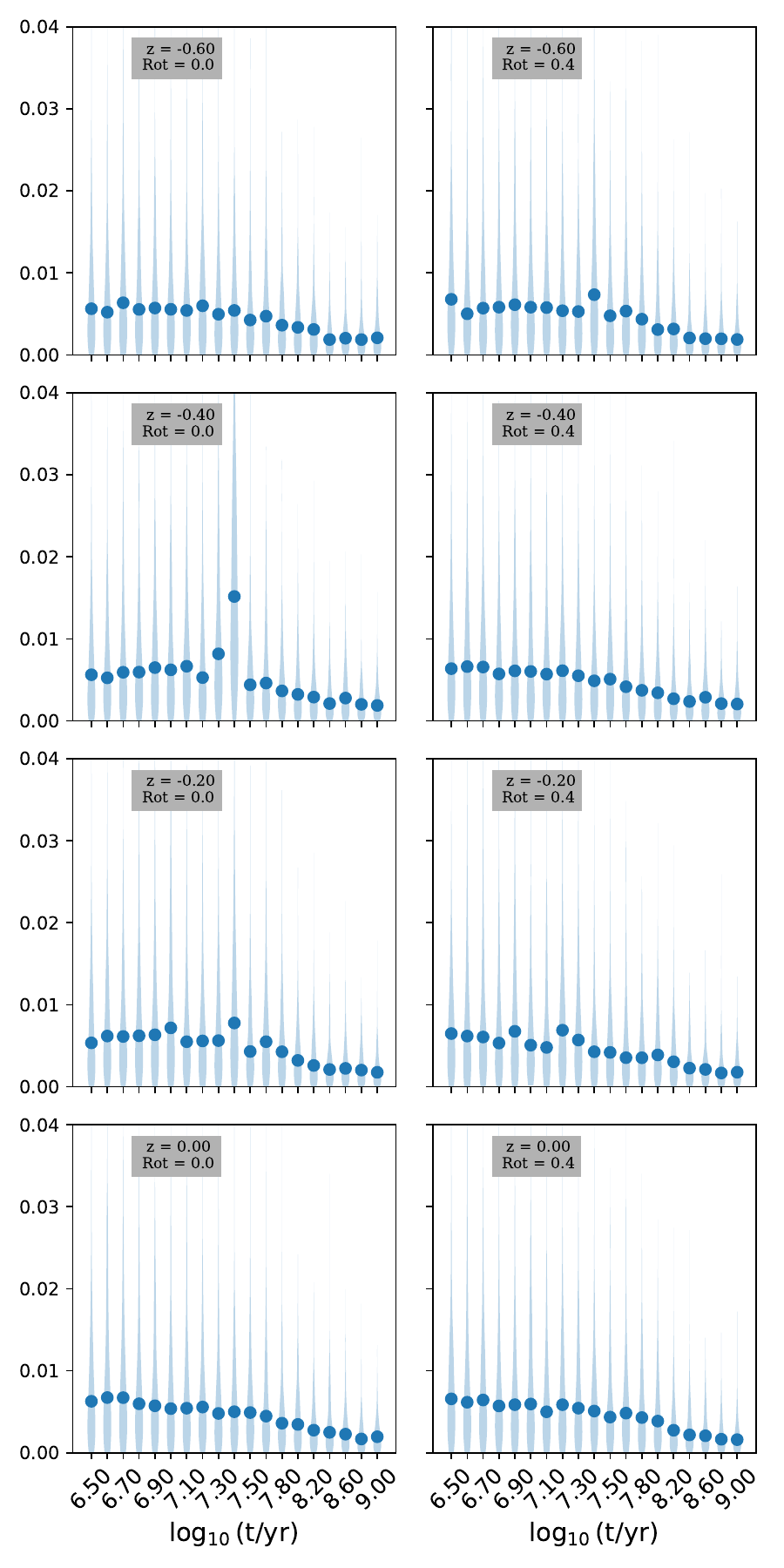}
    \hfill
    \includegraphics[width=0.325\textwidth]{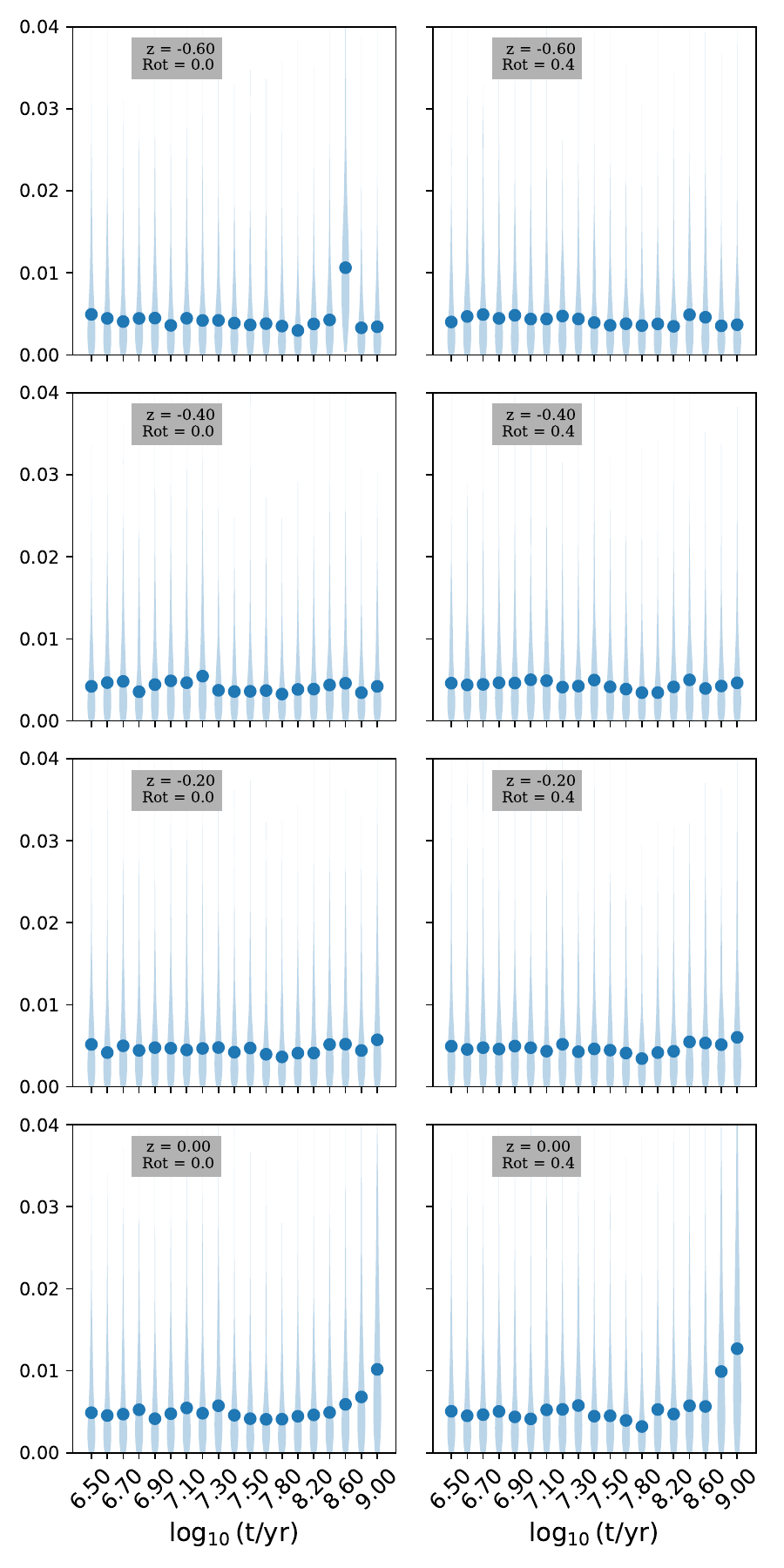}
    \hfill
    \includegraphics[width=0.325\textwidth]{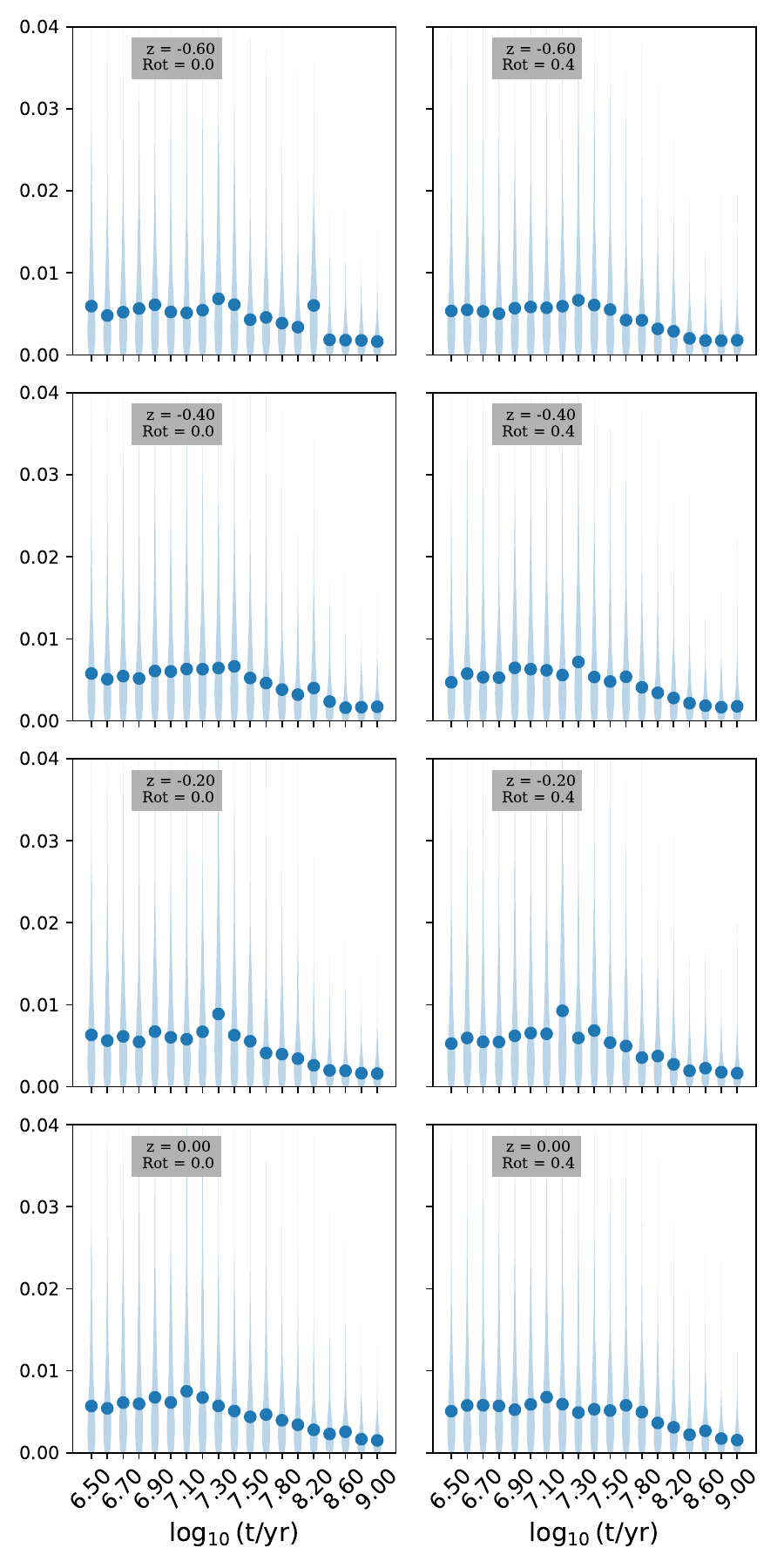}

    
    \makebox[0.325\textwidth][c]{(a)} \hfill
    \makebox[0.325\textwidth][c]{(b)} \hfill
    \makebox[0.325\textwidth][c]{(c)}
    
    \caption{Inferred ages, metallicities, and rotation weights for NGC~2004 (a : left two columns), NGC~7419 (b: middle two columns), and NGC~2100 (c : right two columns), based on MCMC sampling with MIST single-star evolutionary models. Each violin summarizes the posterior distribution of weights, with the central dot representing the median. None of the clusters show a single dominant solution, underscoring the emerging inconsistencies. That said, some of the strongest posterior signals for all three clusters lie near $\log_{10}$(t/yr) $\sim$ 7.3–7.4 ($\sim$20–25 Myr), in good agreement with the RSG-based ages of $\sim$20–24 Myr reported by \citet{Beasor2019}. These preferred solutions contrast with their younger MS and luminosity-function ages of $\sim$7–10 Myr, which we find less support for, suggesting that the stars driving these younger age estimates may not represent the bulk cluster population.}
    \label{fig:clusters_WeightGrid}
\end{figure*}

\begin{figure*}[htbp]
    \centering
    \includegraphics[width=0.325\textwidth]{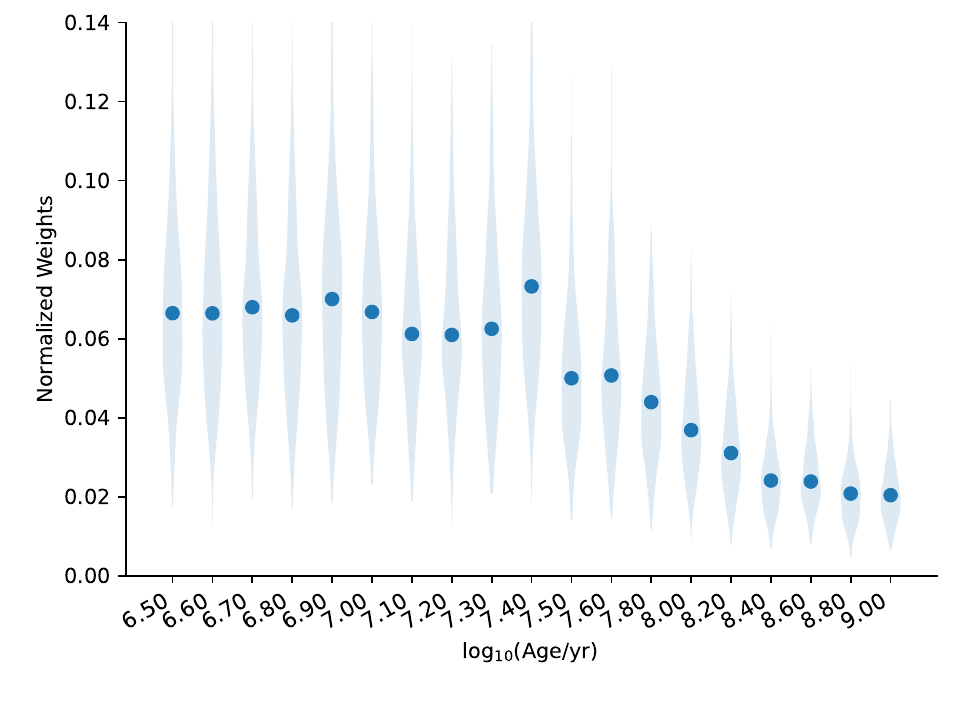}
    \includegraphics[width=0.325\textwidth]{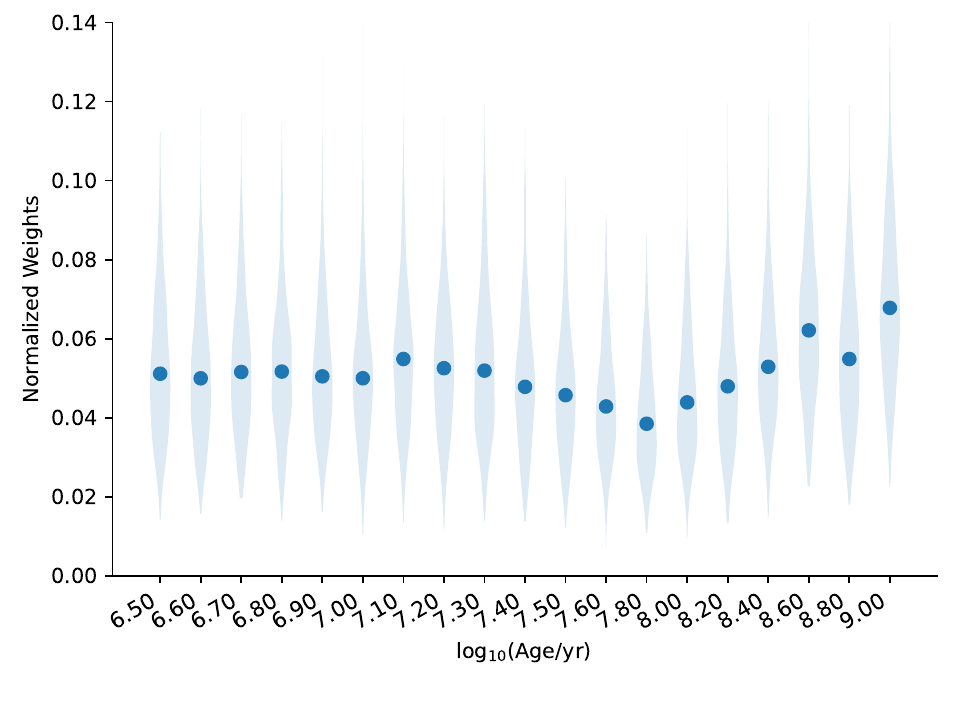}
    \includegraphics[width=0.325\textwidth]{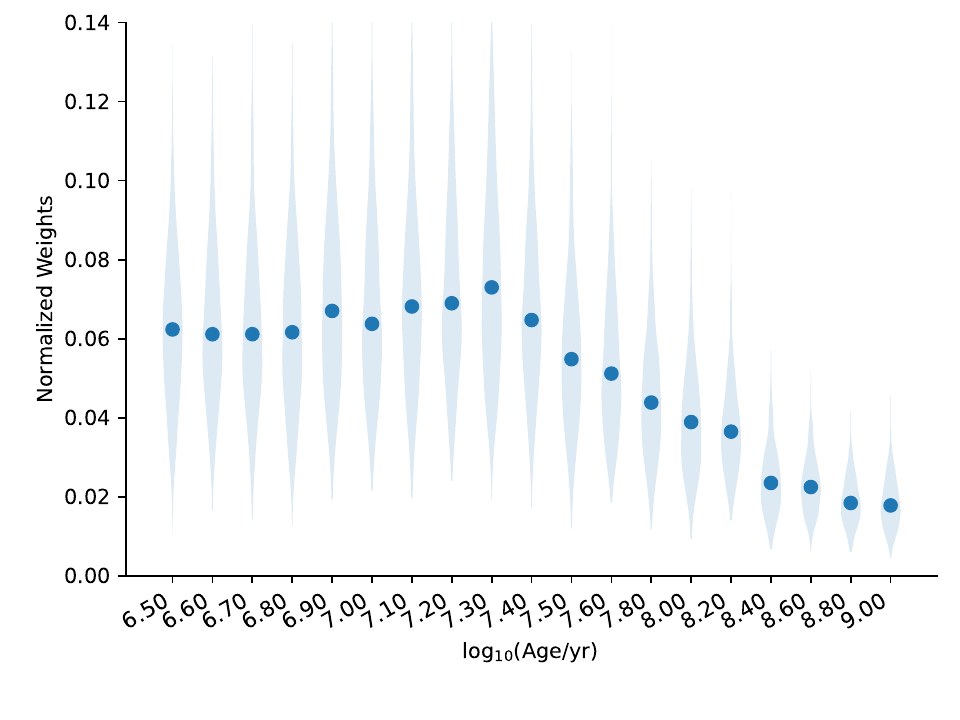}
    
    \vspace{-0.1cm}
    
    \includegraphics[width=0.325\textwidth]{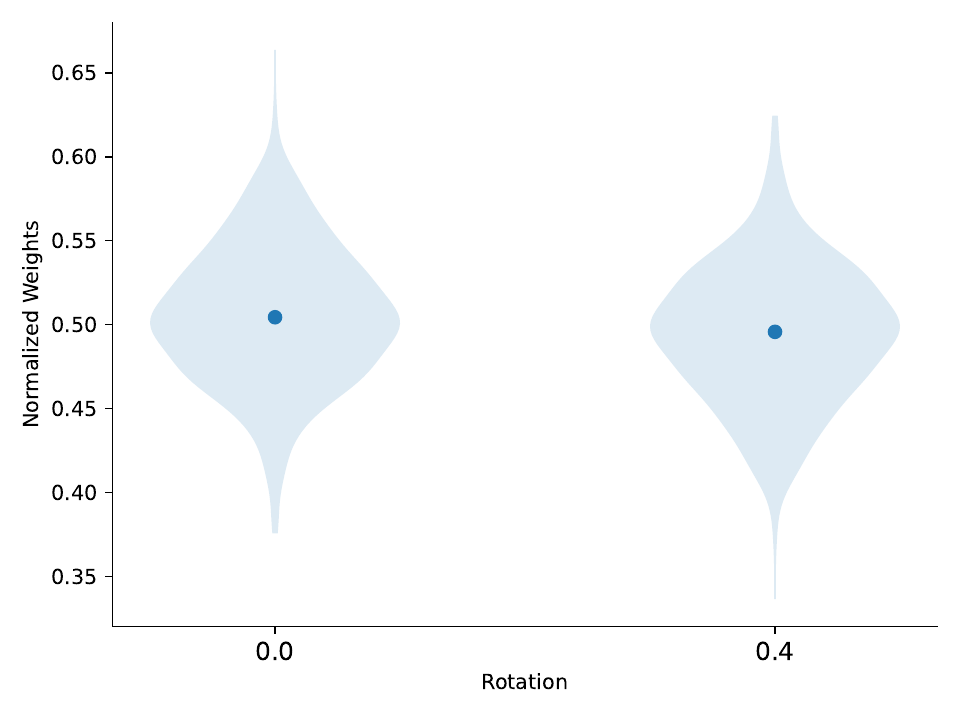}
    \hfill
    \includegraphics[width=0.325\textwidth]{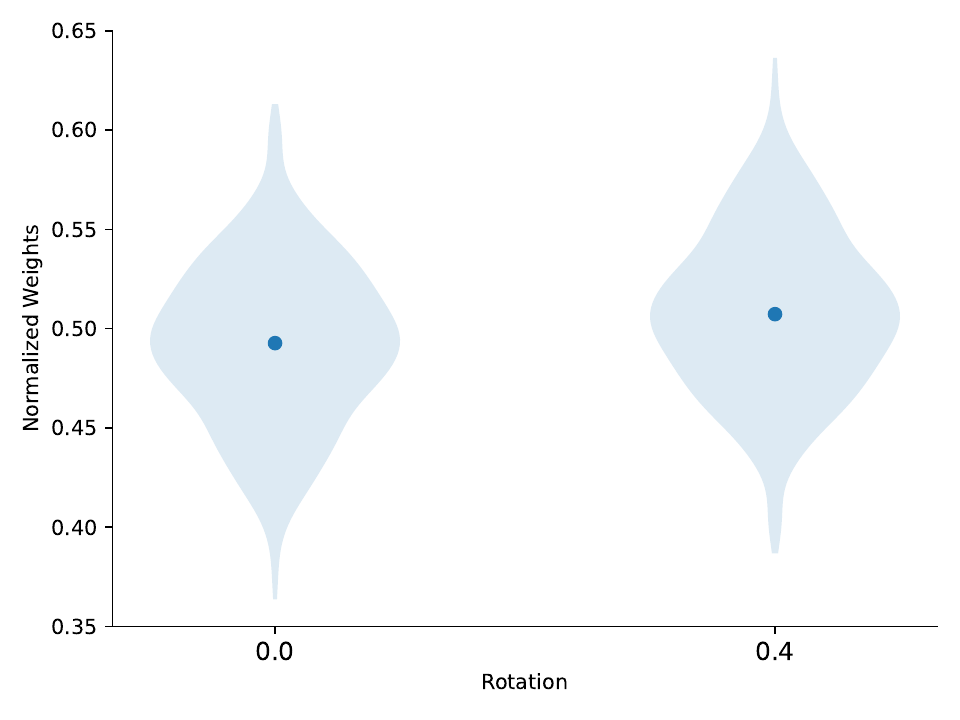}
    \hfill
    \includegraphics[width=0.325\textwidth]{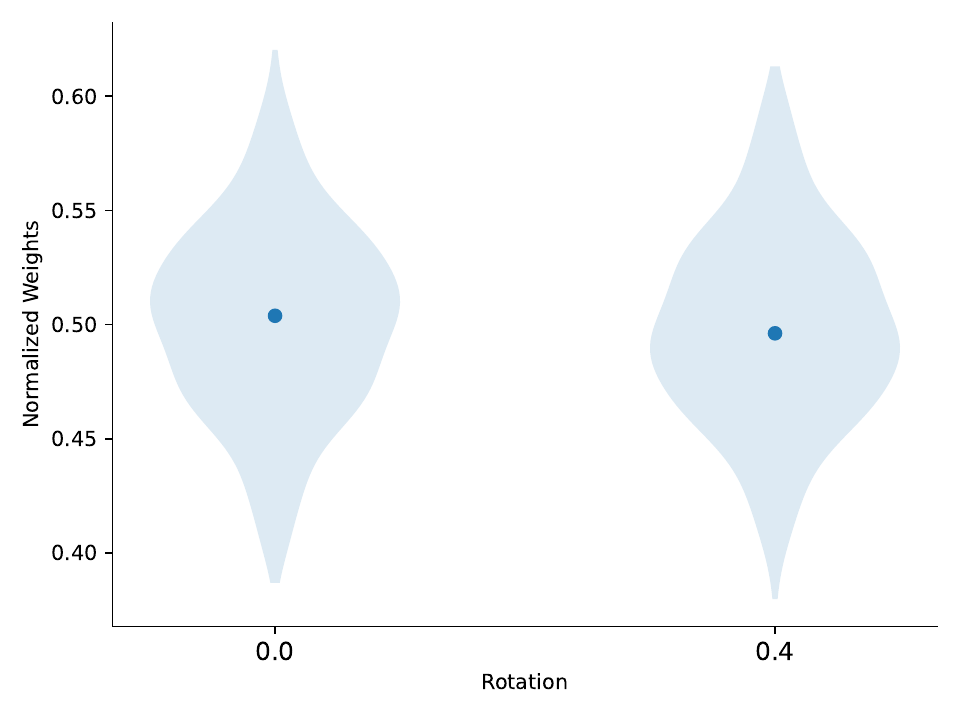}
    
    \makebox[0.325\textwidth][c]{(a)} \hfill
    \makebox[0.325\textwidth][c]{(b)} \hfill
    \makebox[0.325\textwidth][c]{(c)}
    
    \caption{Posterior distributions of age (top row) and initial rotation (bottom row) weights for NGC 2004 (a: left column), NGC 7419 (b: middle column), and NGC 2100 (c: right column), marginalized over the other parameters. The age distributions are broad, with only a modest peak near $\log_{10}$(t/yr) $\sim$ 7.3 ($\sim$20 Myr), and NGC 7419 shows some additional support at older ages. The rotation distributions are flat, with no preference for a specific initial value, indicating that rotation provides little leverage in constraining these clusters. These results are consistent with \citet{Beasor2019}, who found relatively minor shifts in inferred cluster ages when including rotation.}
    \label{fig:clusters_weights}
\end{figure*}

\begin{figure*}[t]
    \centering

    \includegraphics[width=0.48\textwidth]{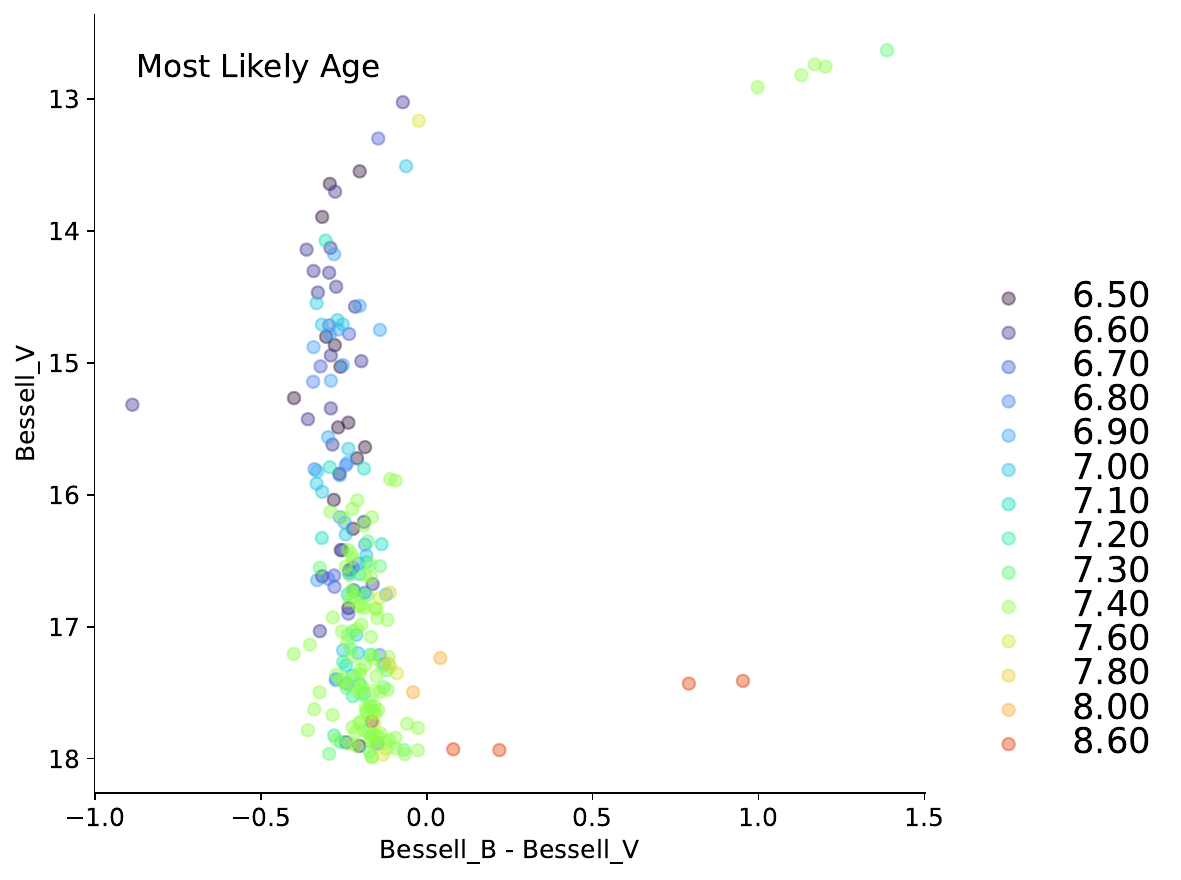}
    \hfill
    \includegraphics[width=0.48\textwidth]{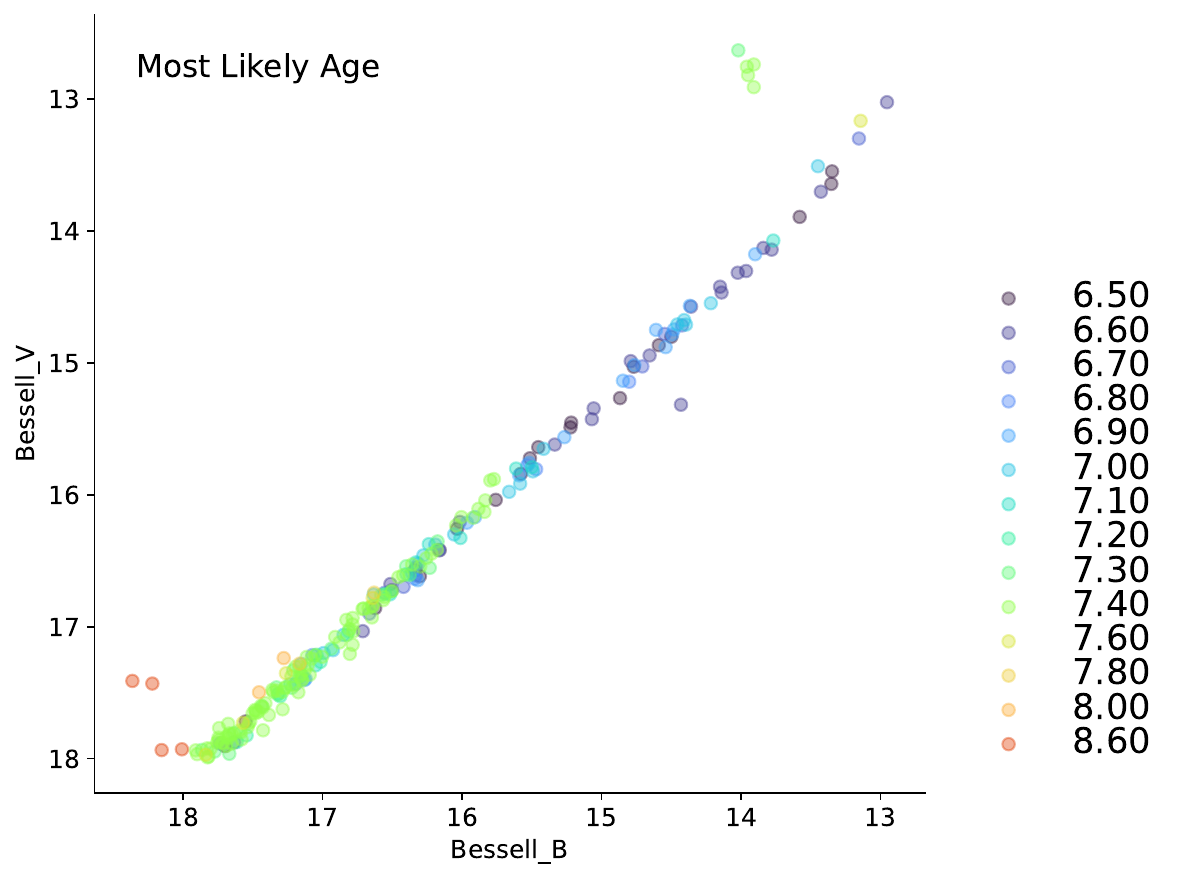}

    \includegraphics[width=0.48\textwidth]{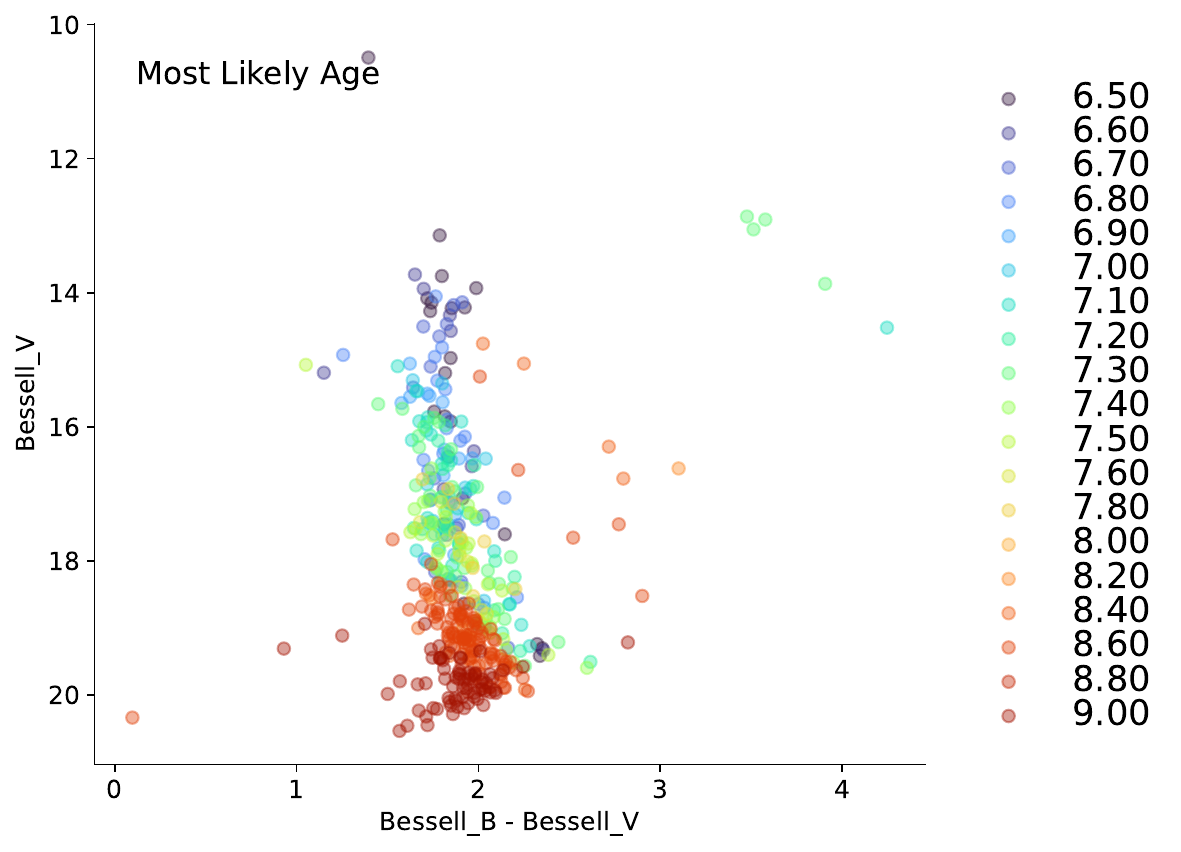}
    \hfill
    \includegraphics[width=0.48\textwidth]{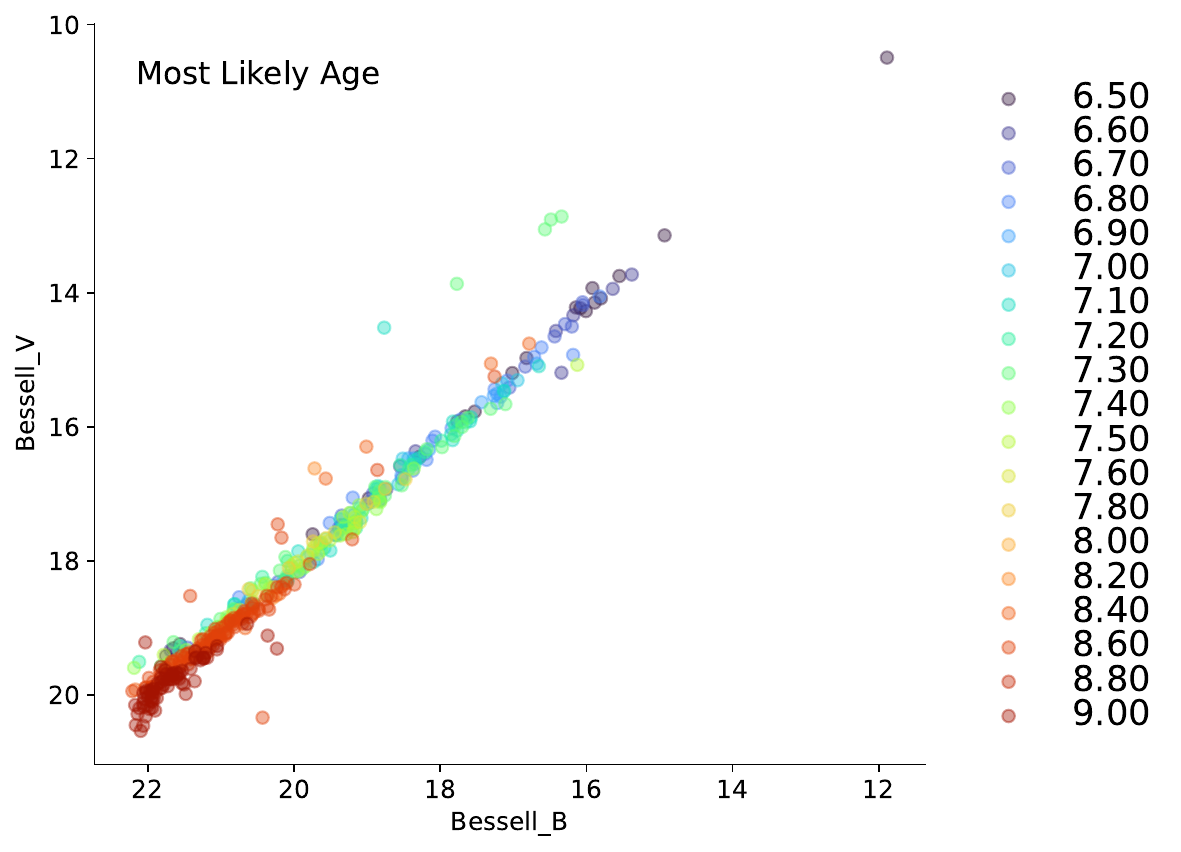}

    \includegraphics[width=0.48\textwidth]{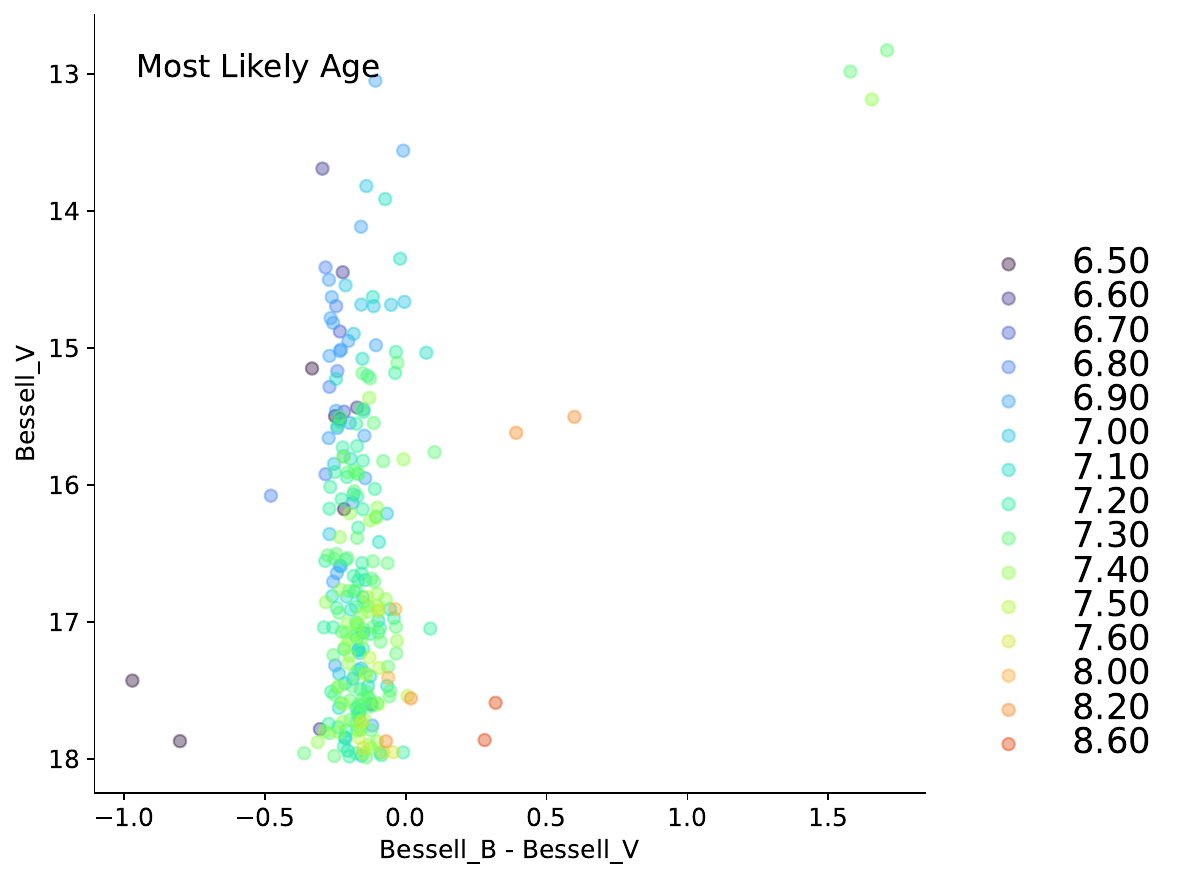}
    \hfill
    \includegraphics[width=0.48\textwidth]{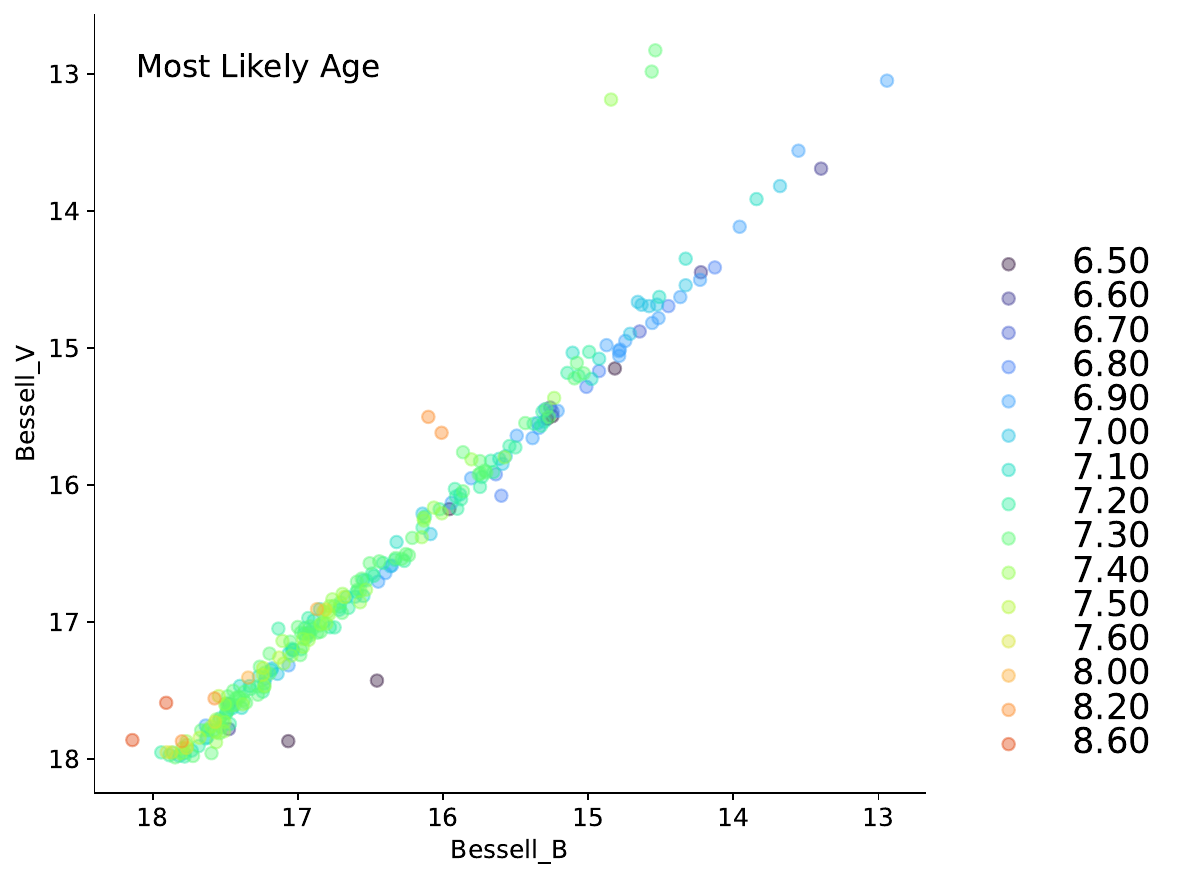}

    \caption{
    Color–magnitude diagrams (left column) and magnitude–magnitude diagrams (right column) with the most likely
    $\log_{10}$ age for each star in NGC~2004 (top row), NGC~7419 (middle row), and NGC~2100 (bottom row).
    Across all three clusters, the $\sim$20 Myr solution (log age $\sim$ 7.3) is supported by the RSGs and a consistent
    main-sequence population, while a younger $\sim$3 Myr signal (log age $\sim$ 6.5) arises from a small number of bright
    blue stars that lack corresponding lower-mass main-sequence members.
    }
    \label{fig:clusters_mostlikely}
\end{figure*}

Figures~\ref{fig:clusters_WeightGrid}–\ref{fig:clusters_mostlikely} present the posterior distributions for age, metallicity, and rotation in four complementary diagnostic views. Figure~\ref{fig:clusters_WeightGrid} summarizes the full three-dimensional posterior distribution of age, metallicity, and rotation. Figure~\ref{fig:clusters_weights} shows the marginalized distributions as a function of age (top row) and rotation (bottom row). Finally, Figure~\ref{fig:clusters_mostlikely} shows the color-magnitude (top row) and magnitude-magnitude  (bottom row) diagrams, and the color of each dot indicates the most likely age.

The full three-dimensional posterior (Figure~\ref{fig:clusters_WeightGrid}) shows that none of the clusters has an overwhelmingly clear age.  For NGC~2004, the strongest weight has an age of $\log_{10}$(t/yr) $\sim$ 7.40, a rotation of 0.0, and a metallicity of [M/H] $\sim$ -0.4. NGC~7419 exhibits multiple peaks corresponding to older ages, $\log_{10}$(t/yr) $\sim$ 8.6, a rotation of 0.0, and a metallicity of [M/H] $\sim$ -0.6, while also showing a young age signal at $\log_{10}$(t/yr) $\sim$ 7.30, rotation 0.0, and [M/H] $\sim$ 0.0. NGC~2100 displays some degeneracy among the parameters, but the clearest signal corresponds to $\log_{10}$(t/yr) $\sim$ 7.30, rotation 0.0, and [M/H] $\sim$ 0.0.  All three clusters are clear spatial concentrations of stars, and as such, they presumably have the same age.  The lack of a clearly statistically significant age is the first sign that there is an internal inconsistency in either the model or data. 

Figure \ref{fig:clusters_weights} shows the marginalized posterior distributions versus~age (top row) and versus~ rotation (bottom row).  The marginalized distributions versus~age show broad profiles with a modest bump near $\log_{10}$(age/yr) $\approx$ 7.3, a feature that will gain additional context in later analysis. However, as in the full 3D posterior distributions, there are no clear statistically significant ages for any of the clusters. Notably, NGC~7419 shows a weak signal also for older ages.

The marginalized distributions as a function of rotation show that there is no preference for a particular rotation, indicating that the initial rotation values provide little discriminatory power for distinguishing cluster-specific formation conditions or evolutionary histories. This corresponds with the results of \citet{Beasor2019}, which indicated that the inclusion of stellar models with rotation led to relatively minor differences in the inferred age of these clusters. 

These distributions for the weights represent the entire stellar population, which can obscure features that are particularly relevant for specific evolutionary stages, such as the bright evolved stars that dominate the pertinent age signals.

We now turn from population-level weights to the inferences for each individual star provided by Stellar Ages.  Figure~\ref{fig:clusters_mostlikely} shows the most likely age on CMDs (top row) and magnitude-magnitude diagrams (bottom row).  This added level of inference provides greater clarity on the lack of strong age probabilities. 

Our analysis reveals several compelling patterns across all three clusters. In general, the RSGs have ages around $\log_{10}$(t/yr) $\approx$ 7.30 (20 Myr), with corresponding MS stars that support this age. However, we identify a population of bright blue stars in all three clusters that suggest significantly younger ages of $\log_{10}$(t/yr)$\sim$6.5 (3.2 Myr). Interestingly, these young stars appear as outliers in the CMD, lacking the expected population of lower-mass main sequence stars under a standard IMF. Some likely resolutions to this inconsistency are that either the bright blue stars are the product of binary evolution, they are rapid rotators, or the bright blue stars are a combination of both. See \citet{Murphy_2025} for a more thorough discussion on these and other possible resolutions to this inconsistency.

A population of bright blue stars appearing younger than expected naturally brings to mind blue stragglers or an extended MSTO (eMSTO). As we discuss in Section~\ref{subsec:implications}, while these concepts are closely related, the qualitative and quantitative appearance of the bright blue stars here is distinct from both the classic blue straggler phenomenon in old clusters and the eMSTOs seen in intermediate-age systems.

\subsection{Quantifying the Inconsistency Among the Stellar Age Tracers}
\label{subsec:RSGs}

\begin{figure*}[htbp]
    \centering
    
    \begin{minipage}[t]{0.48\textwidth}
        \begin{overpic}[width=\textwidth]{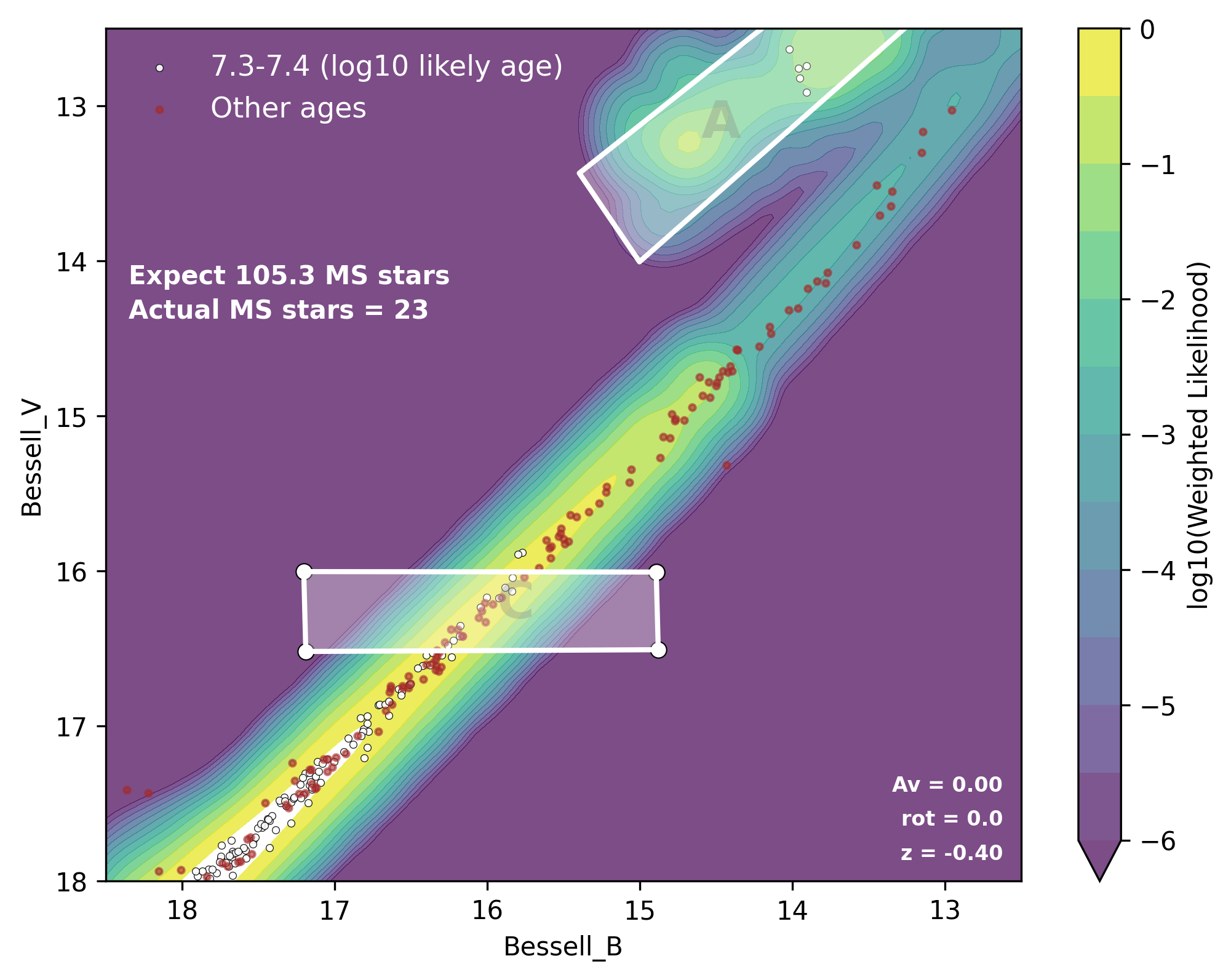}
            \put(35, 12){\textcolor{white}{\textbf{NGC 2004}}}
        \end{overpic}
    \end{minipage}
    \hfill
    \begin{minipage}[t]{0.48\textwidth}
        \begin{overpic}[width=\textwidth]{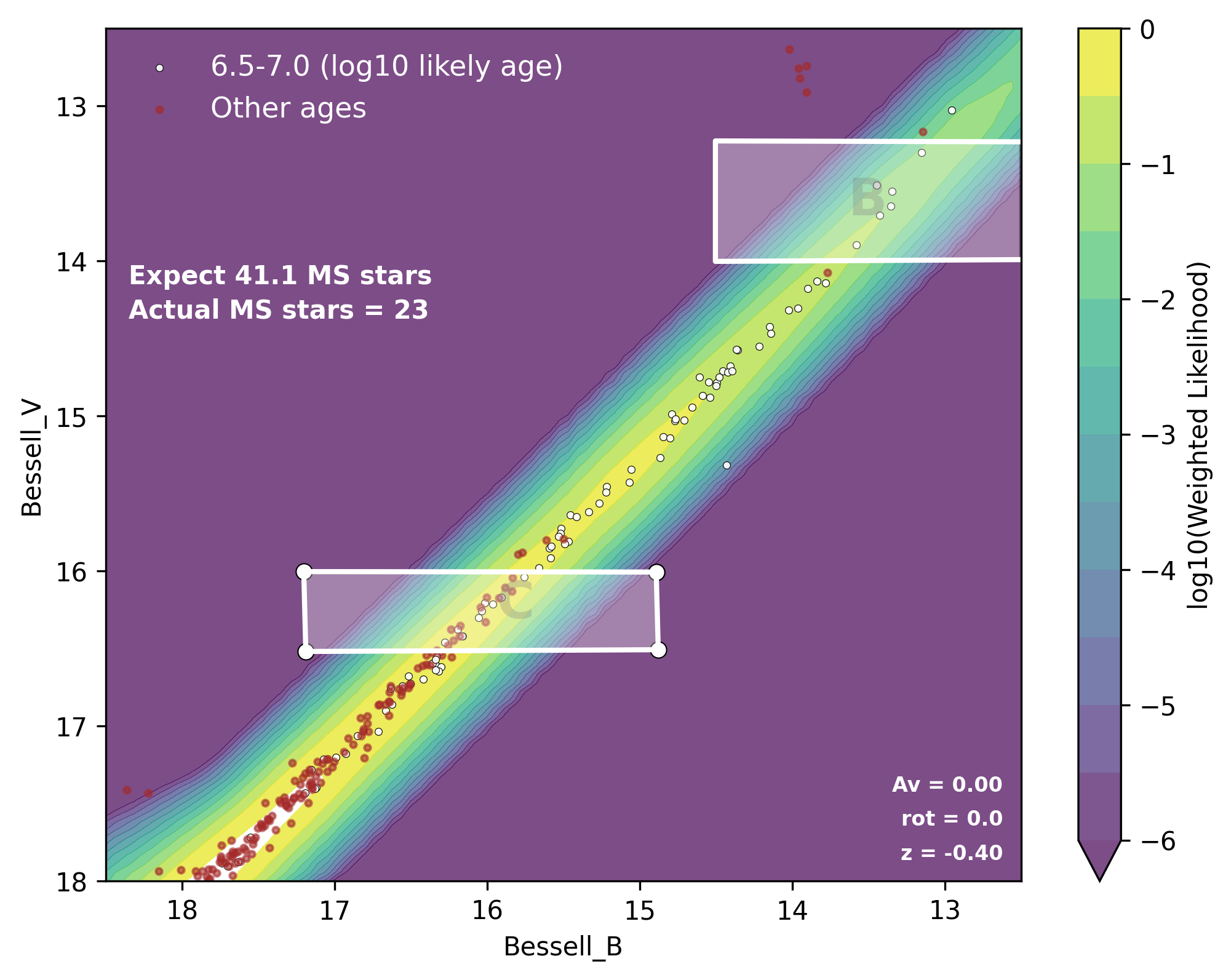}
            \put(35, 12){\textcolor{white}{\textbf{NGC 2004}}}
        \end{overpic}
    \end{minipage}
    
    \begin{minipage}[t]{0.48\textwidth}
        \begin{overpic}[width=\textwidth]{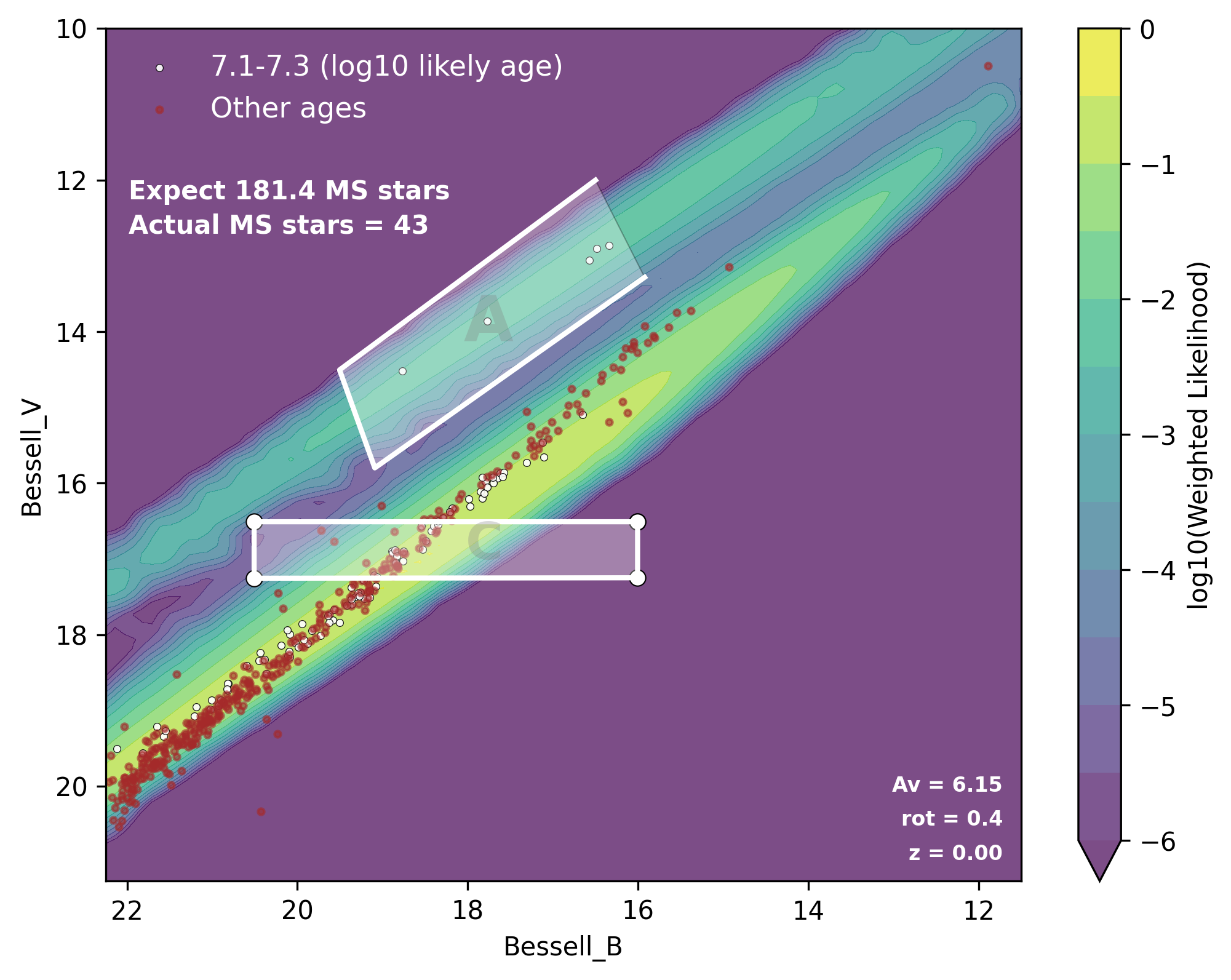}
            \put(35, 12){\textcolor{white}{\textbf{NGC 7419}}}
        \end{overpic}
    \end{minipage}
    \hfill
    \begin{minipage}[t]{0.48\textwidth}
        \begin{overpic}[width=\textwidth]{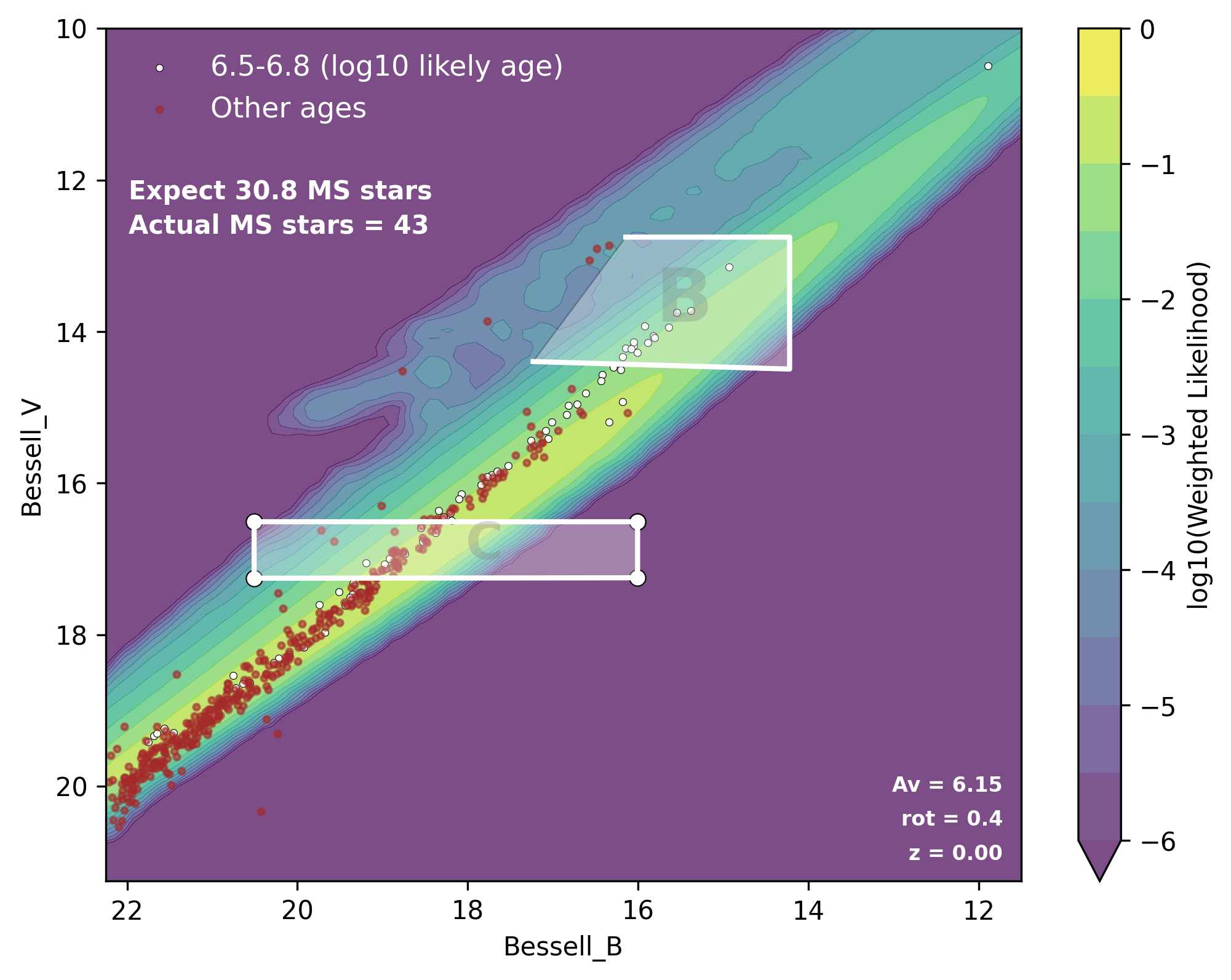}
            \put(35, 12){\textcolor{white}{\textbf{NGC 7419}}}
        \end{overpic}
    \end{minipage}
    
    \begin{minipage}[t]{0.48\textwidth}
        \begin{overpic}[width=\textwidth]{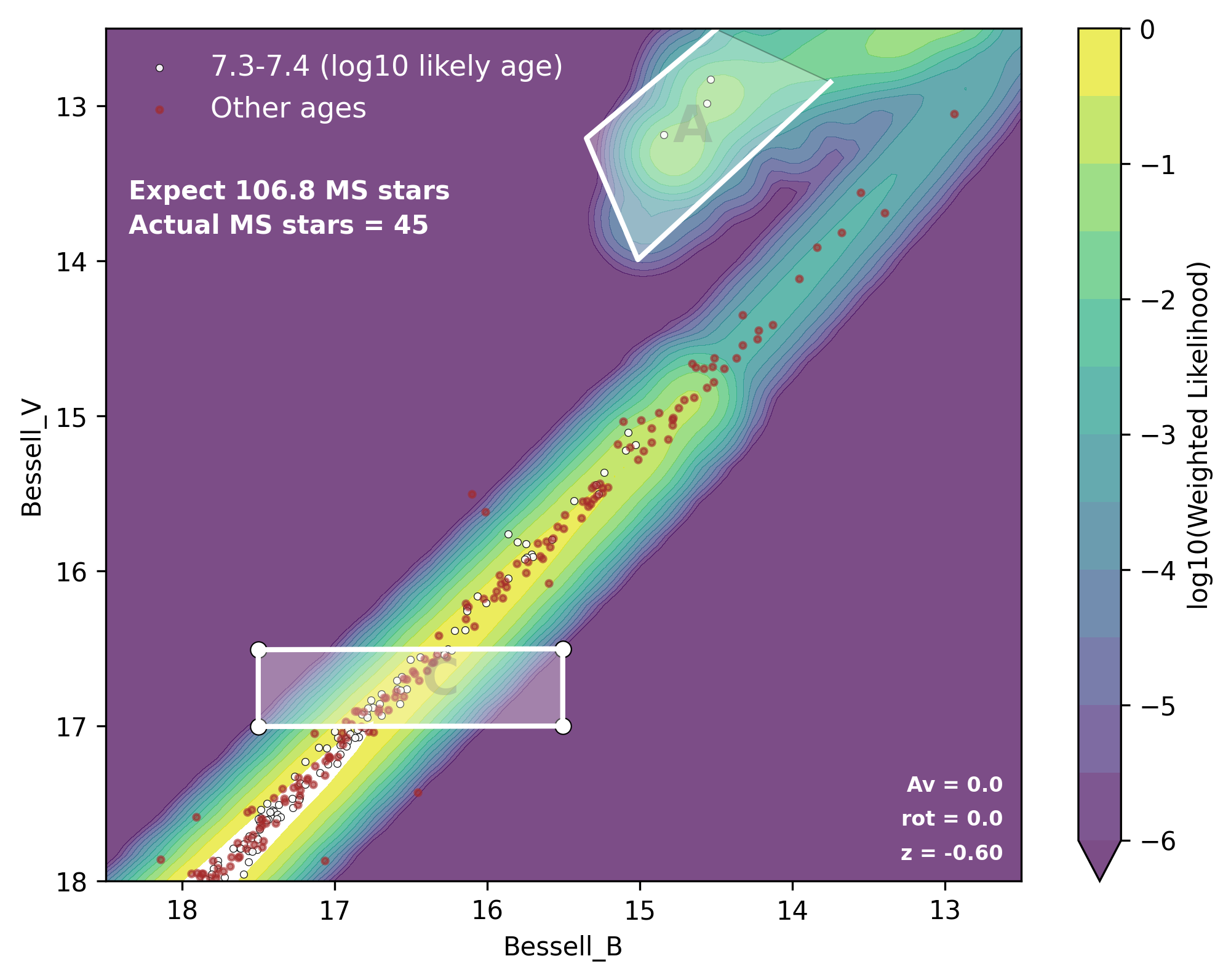}
            \put(35, 12){\textcolor{white}{\textbf{NGC 2100}}}
        \end{overpic}
    \end{minipage}
    \hfill
    \begin{minipage}[t]{0.48\textwidth}
        \begin{overpic}[width=\textwidth]{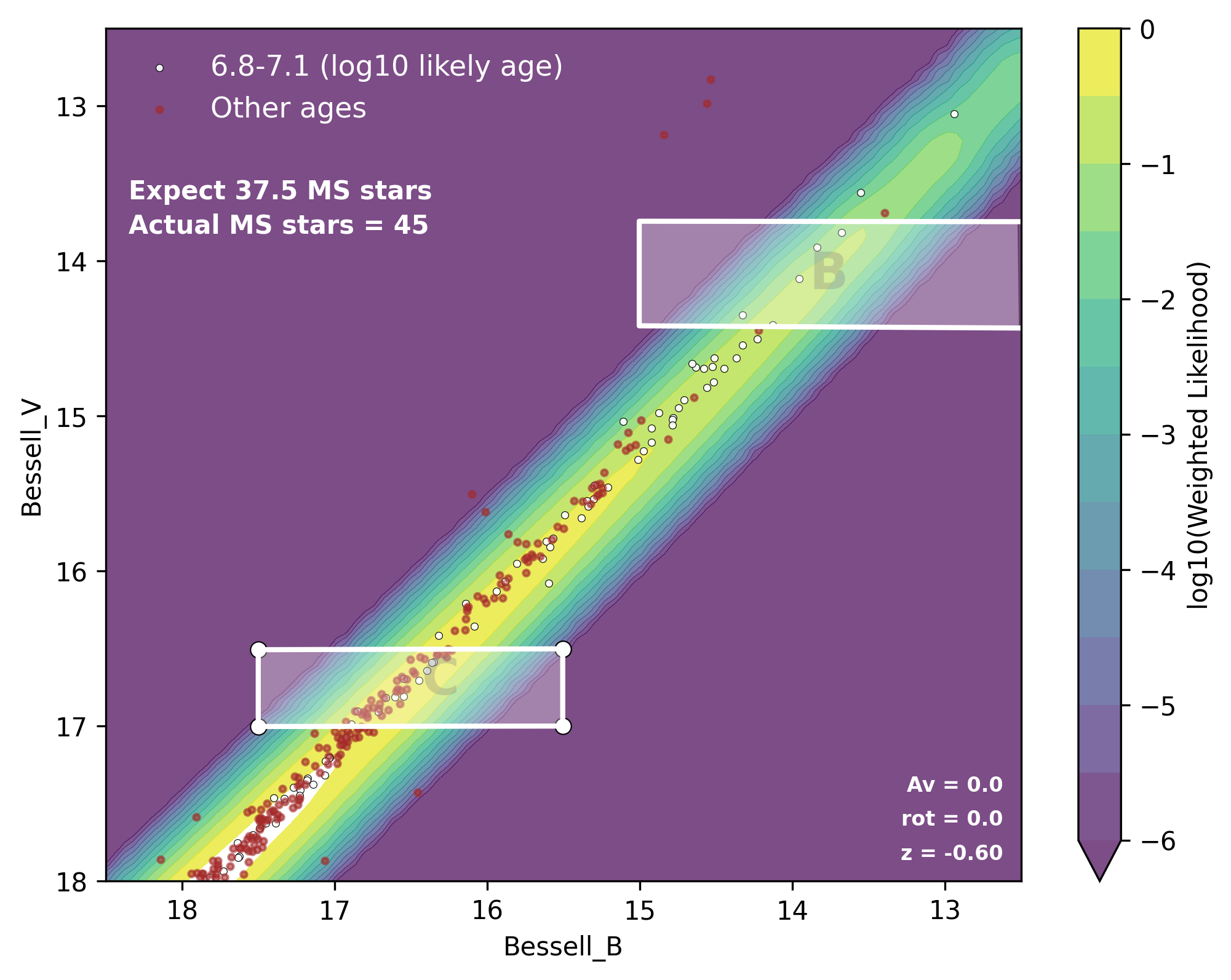}
            \put(35, 12){\textcolor{white}{\textbf{NGC 2100}}}
        \end{overpic}
    \end{minipage}
    
    \caption{These panels compare the expected MS counts implied by evolved stars to the observed counts. The rows separate the three clusters, top: NGC~2004, middle: NGC~7419, bottom: NGC~2100. Left column uses RSGs (region A) as predictors and right column uses bright blue stars (region B). The background shows the joint age probability density functions inferred for the predictor region. White points mark stars whose most likely ages fall within the inferred age range of the predictor region. Brown points mark other stars. Region C is the common MS comparison region. Expected counts are computed with the weighted model described in subsection \ref{subsec:RSGs}}
    \label{fig:clusters_expectednum}
\end{figure*}

To quantify the inconsistency identified in the previous section, Figure~\ref{fig:clusters_expectednum} compares the expected number of stars in various parts of the CMD with the actual data. For each cluster, we defined three regions on the diagrams: Region A capturing RSGs (left column), Region B containing bright blue stars (right column), and Region C is a consistent main sequence (MS) comparison region. We then calculated the expected number of MS stars given the number of observed evolved stars in Regions A or B and compared this with the actual MS count. This statistical test provides a quantitative measure of population consistency and highlights implications for the underlying cluster structure, similar to the analysis presented in \cite{Murphy_2025}.

\begin{figure}[htbp]
    \centering
    \begin{overpic}[width=0.85\columnwidth]{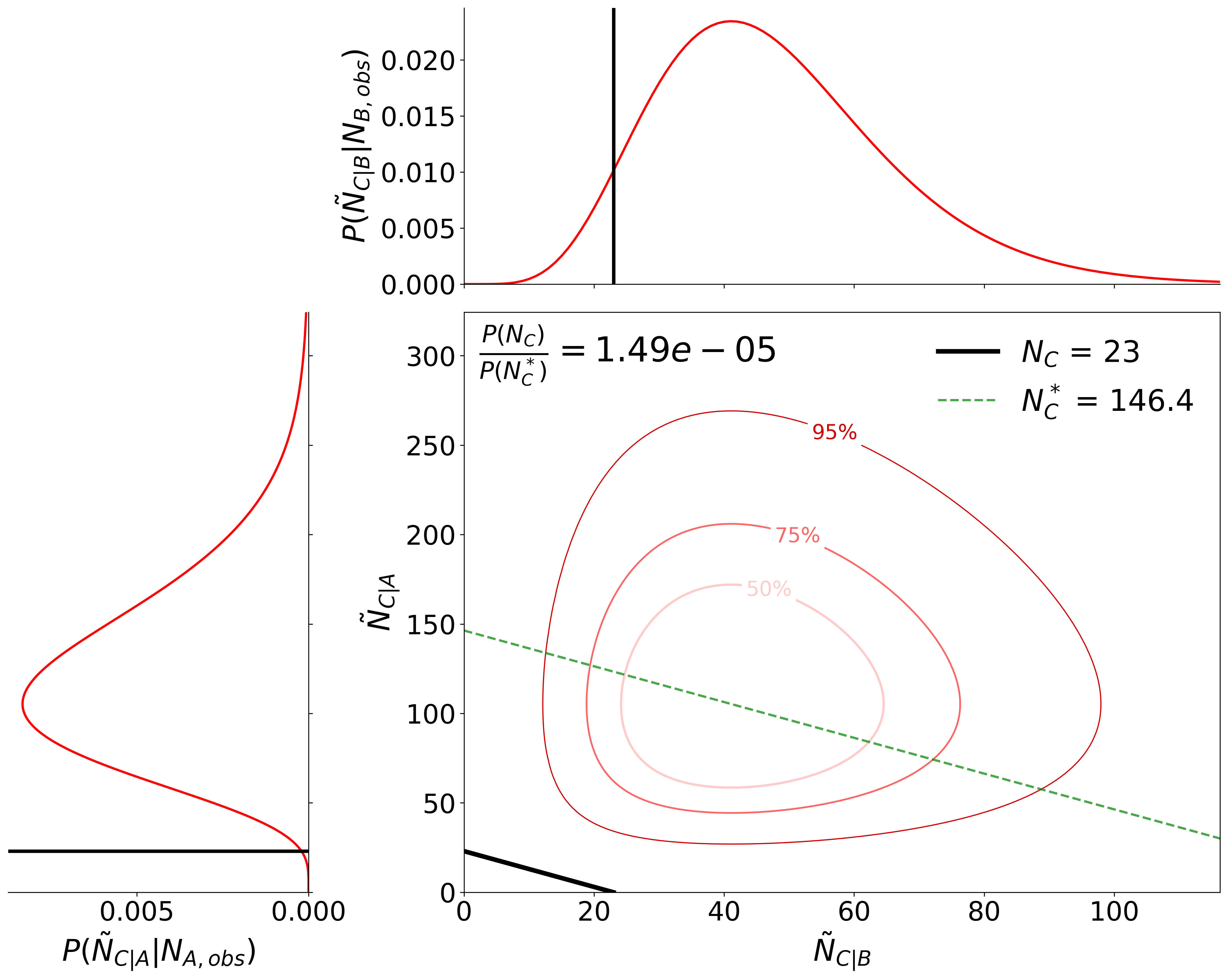}
        \put(12,65){\textcolor{black}{\textbf{(a)}}}
    \end{overpic}
    
    \vspace{0.05cm}
    
    \begin{overpic}[width=0.85\columnwidth]{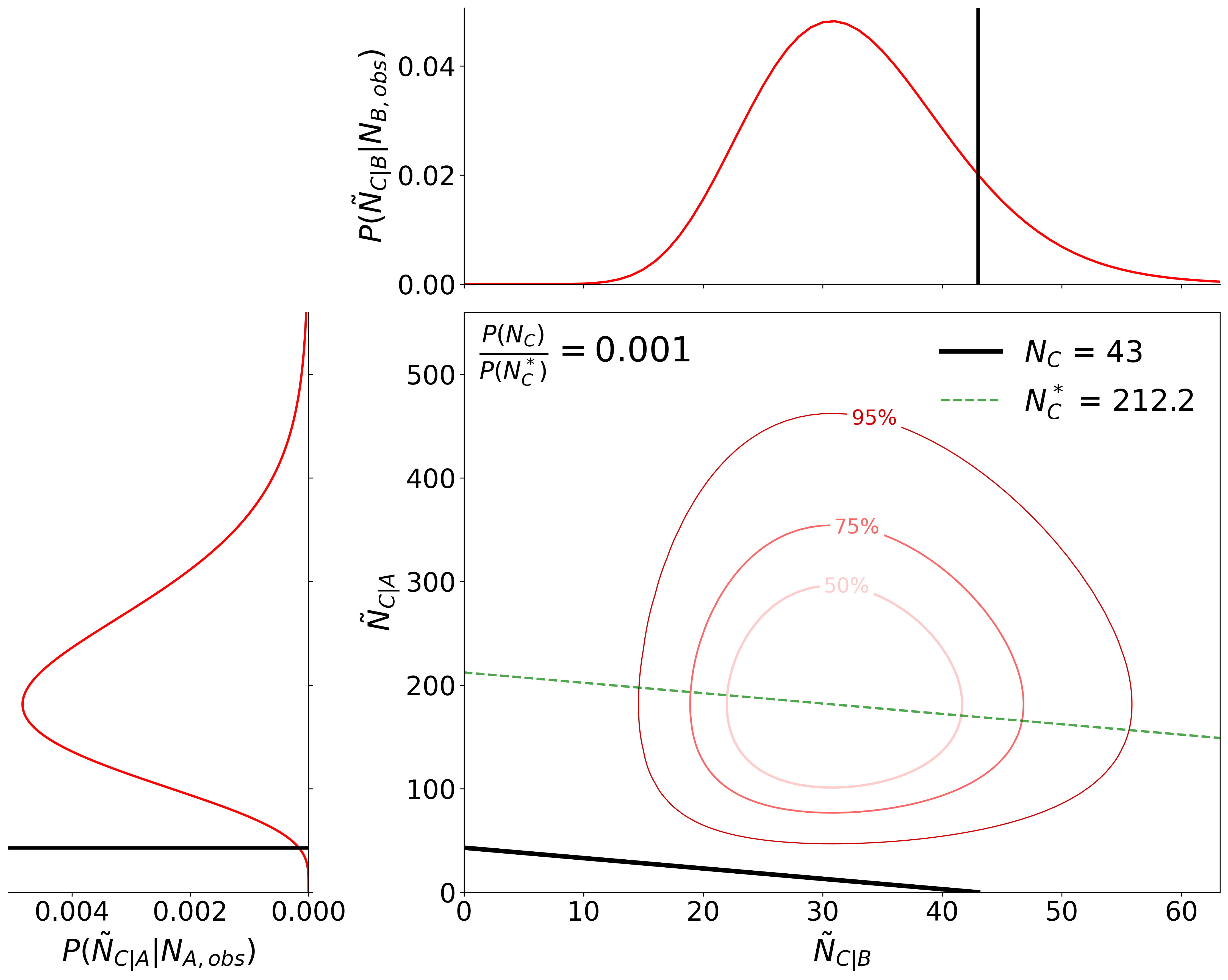}
        \put(12,65){\textcolor{black}{\textbf{(b)}}}
    \end{overpic}
    
    \vspace{0.05cm}
    
    \begin{overpic}[width=0.85\columnwidth]{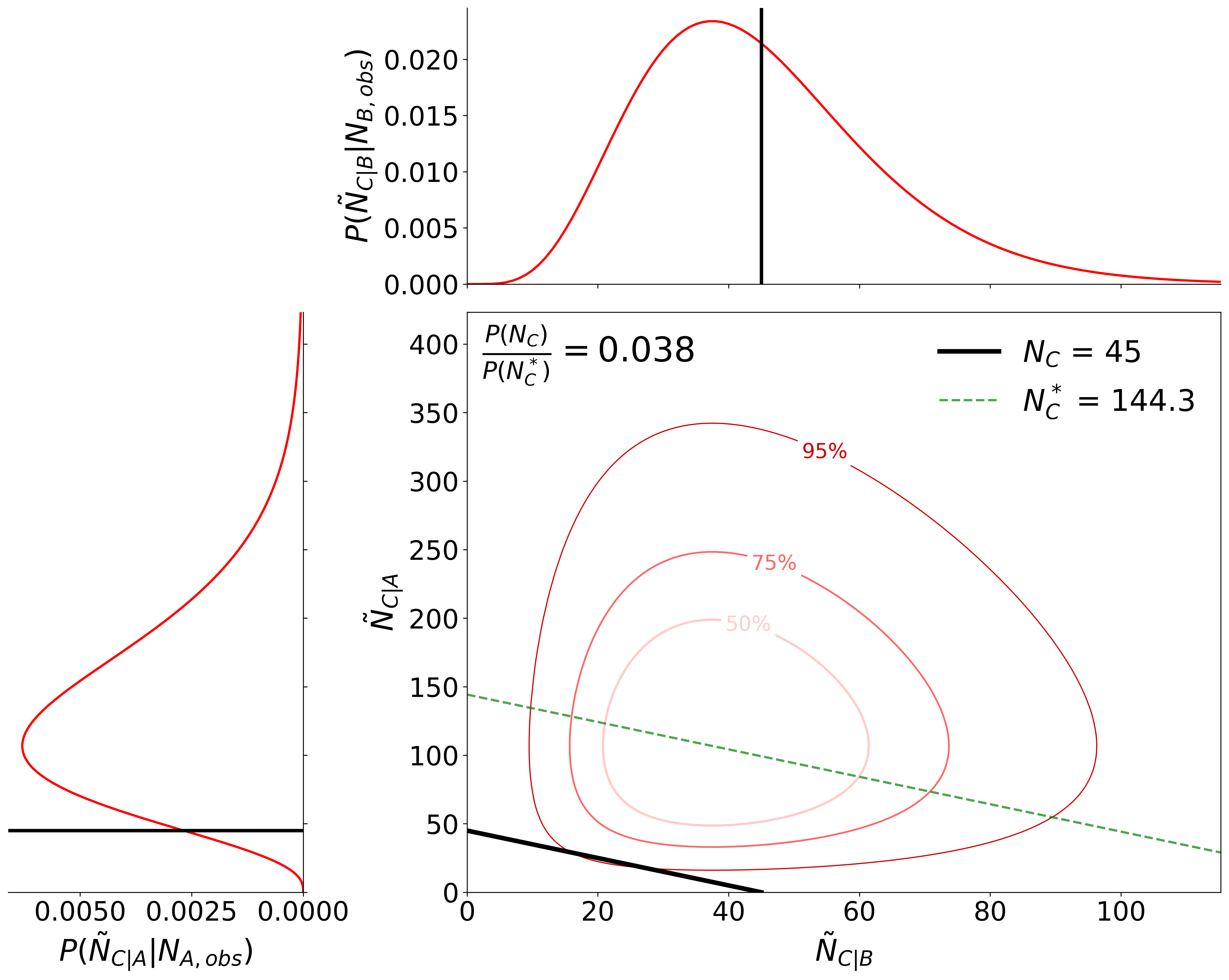}
        \put(12,65){\textcolor{black}{\textbf{(c)}}}
    \end{overpic}
    
    \caption{Posterior distribution for the expected number of MS stars in Region C predicted from the evolved populations for NGC 2004 (a: top panel), NGC 7419 (b: middle panel), and NGC 2100 (c: bottom panel). The parameters $\tilde{N}_{C|A}$ and $\tilde{N}_{C|B}$ represent the expected contributions from RSGs (Region A) and bright blue stars (Region B), respectively. The red contours show the joint posterior distribution with 50\%, 75\%, and 95\% highest-density credible intervals (HDCIs). The solid black line marks the observation constraint, $\tilde{N}_{C|A} + \tilde{N}_{C|B} = N_{C,obs}$, while the green dashed line marks the model-consistent expectation, $\tilde{N}_{C|A} + \tilde{N}_{C|B} = N_C^*$. The marginalized posteriors along the top and left axes illustrate the distributions for $\tilde{N}_{C|A}$ and $\tilde{N}_{C|B}$ individually. The strong offset between the observed and model-consistent constraint lines highlights the systematic tension between predictions and the observed MS population.}
    \label{fig:clusters_posterior}
\end{figure}

This framework yields separate expectations for the MS implied by RSGs and by bright blue stars, which we compare directly to the observed main-sequence counts. The expected number of MS stars from each evolved population is obtained from a likelihood ratio. For a given evolved population with $N_{\rm evolved}$ stars, the expected MS stars $N_{\rm MS,expected}$ is determined by:
\begin{equation}
N_{\rm MS,expected} = N_{\rm evolved} \times \frac{\int_{\rm MS} \mathcal{L}(\theta) \, d\theta}{\int_{\rm evolved} \mathcal{L}(\theta) \, d\theta}
\end{equation}
where $\mathcal{L}(\theta)$ represents the likelihood function over the parameter space ($\theta = \{t, \rm{[M/H]}, v_{\rm{ini}} \}$), and the integrals are taken over the MS and evolved star region (region A or region B), respectively. This ratio quantifies the relative probability of finding stars in the MS region versus the evolved star region based on stellar evolution models, convolved with the IMF, and the observation model.

To assess the statistical significance of any discrepancy between expected and observed MS populations, we calculate the posterior of the expected number of MS stars (region C) given the number of RSGs (region A) and bright blue stars (region B). The expected number of MS stars in region C given the number RSGs in region A is $\tilde{N}_{C|A}$ and the equivalent given bright blue stars in region B is $\tilde{N}_{C|B}$. To construct a joint posterior distribution for these expected numbers, we first construct a joint likelihood for the observing $N_{A,{\rm obs}}$ RSGs and $N_{B,{\rm obs}}$ bright blue stars, where the likelihood model is the Poisson distribution:
\begin{equation}
\begin{aligned}
P(N_{A,obs} \mid \lambda_A) &= \frac{\lambda_A^{N_{A,obs}} e^{-\lambda_A}}{N_{A,obs}!}, \\
P(N_{B,obs} \mid \lambda_B) &= \frac{\lambda_B^{N_{B,obs}} e^{-\lambda_B}}{N_{B,obs}!}.
\end{aligned}
\end{equation}

\noindent The expected number of stars in this region $\lambda$ is 

\begin{equation}
\lambda_A = \frac{\tilde{N}_{C|A}}{r_A}, \quad \lambda_B = \frac{\tilde{N}_{C|B}}{r_B}
\end{equation}

\noindent where $r_A$ and $r_B$ are the likelihood ratios between MS and evolved star regions.
The joint posterior is then:
\begin{equation}
\begin{aligned}
P(\tilde{N}_{C|A}, \tilde{N}_{C|B} \mid N_{A,obs}, N_{B,obs}) 
&\propto P(N_{A,obs} \mid \lambda_A) \\
&\quad \times P(N_{B,obs} \mid \lambda_B).
\end{aligned}
\end{equation}

Figure~\ref{fig:clusters_posterior} shows the posterior distribution for the expected number of MS stars in region C. For each subplot, the marginalized distributions are shown in the left and top panels.  

The expected number of stars from the RSGs and from the bright blue stars must equal the total number of actual observed stars, i.e.
\begin{equation}
\tilde{N}_{C|A} + \tilde{N}_{C|B} = N_{C,obs}
\end{equation}
This constraint is represented by the black dashed line in each panel of Figure~\ref{fig:clusters_posterior}. 

In all three cases, the constraint indicated by the black line lies well outside the bulk of the posterior. If the data were consistent with the standard assumptions in modeling these young stellar populations, then the total number of stars in region C would be
\begin{equation}
\tilde{N}_{C|A} + \tilde{N}_{C|B} = N_C^*
\end{equation}
where $N_C^* = N_{A,obs} \times r_A + N_{B,obs} \times r_B$ is the most theoretically consistent total MS star count. The green line shows the constraint if the actual data were most consistent with model. We label each line with the associated values of $N_{C,obs}$ and $N_{C}^{*}$.

To quantify the statistical tension between the observed and model-consistent scenarios, we compute a support ratio that measures the relative posterior probability along these constraint lines and is displayed as an annotation on the posterior plots:

\begin{equation}
R = \frac{P(\tilde{N}_{C|A} + \tilde{N}_{C|B} = N_{C,obs})}{P(\tilde{N}_{C|A} + \tilde{N}_{C|B} = N_C^*)}
\end{equation}

This ratio compares posterior support along the observed constraint line to the model-consistent line. Values below one indicate the observed line is less supported than the model-consistent line, suggesting potential tension between the data and stellar evolution models.

The construction of this statistical test required careful consideration of several methodological choices to ensure physically meaningful and informative results. Since our analysis marginalizes across age likelihoods, we selected the most probable metallicity and rotation parameters for each cluster based on the posterior distributions shown in Figure \ref{fig:clusters_WeightGrid}. The definition of the three analysis regions (A, B, and C) involved deliberate choices to capture distinct evolutionary phases while maintaining statistical robustness. Region A encompasses the available RSGs along with their surrounding likelihood-weighted areas, ensuring comprehensive sampling of the evolved stellar population. Region B targets a subset of the brightest blue stars, which exhibit characteristic young ages of $\log_{10}$(t/yr) $\sim$ 6.5. Region C was strategically positioned along the main sequence to provide a consistent comparison baseline across the full most likely age range spanned by regions A and B, while carefully avoiding the main-sequence turnoff regions to prevent contamination from either subpopulation, while remaining sufficiently bright to minimize completeness-related biases. Further consideration is needed toward the proper selection of various regions for statistical comparison, however that is an effort that goes beyond the scope of this paper.

Our analysis reveals a striking systematic pattern across all three clusters. RSG populations consistently overpredict MS star counts by substantial factors:  NGC 2004 shows a 4.6 times overprediction (105.3 expected versus 23 observed), NGC 7419 shows a 4.2 times overprediction (181.4 expected versus 43 observed) and NGC 2100 shows a 2.4 times overprediction (106.8 expected versus 45 observed). In contrast, predictions based on bright blue star populations also deviate from the observed MS counts but show better agreement than those inferred from RSGs, with expected-to-observed number ratios ranging from 0.7 to 1.8.

The statistical significance of these discrepancies is quantified by extremely low support ratios when considering both subpopulations: $R = 1.49 \times 10^{-5}$ for NGC 2004, $R = 0.001$ for NGC 7419 and $R = 0.038$ for NGC 2100. These values indicate that the observed MS populations are highly unlikely under standard single-star stellar evolution models, with support ratios deviating far from the model-consistent expectation. The consistency of this pattern across three independent clusters strongly argues against random statistical fluctuations.

In all three clusters, there is a systematic discrepancy between the observed number of RSGs and the expected number of MS stars predicted by single-star models. We will discuss possible resolutions in section \ref{subsec:implications}.

\subsection{Inferring Population Parameters from RSG and Blue Star Subsets}
\label{subsec: siblings}

\begin{figure*}[htbp]
    \centering

    \includegraphics[width=0.83\textwidth]{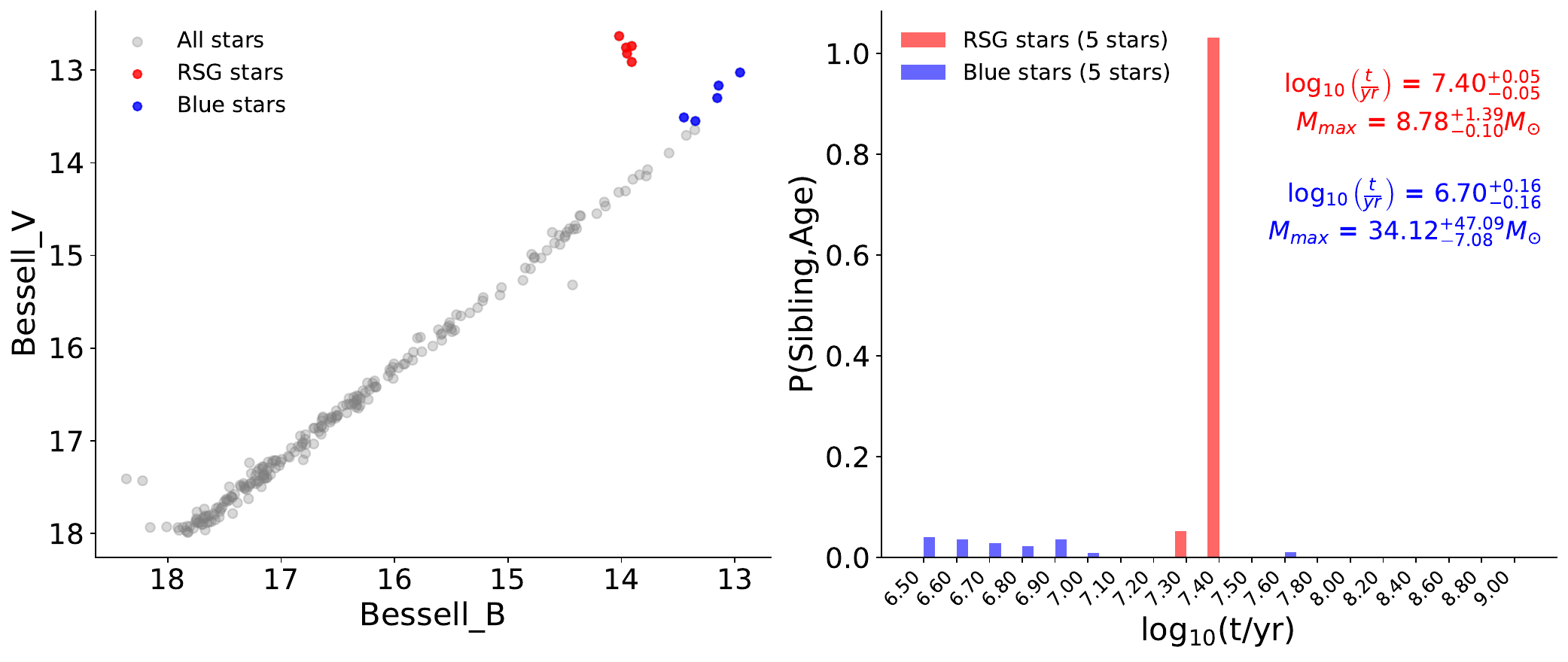}
    \vspace{0.3cm}

    \includegraphics[width=0.83\textwidth]{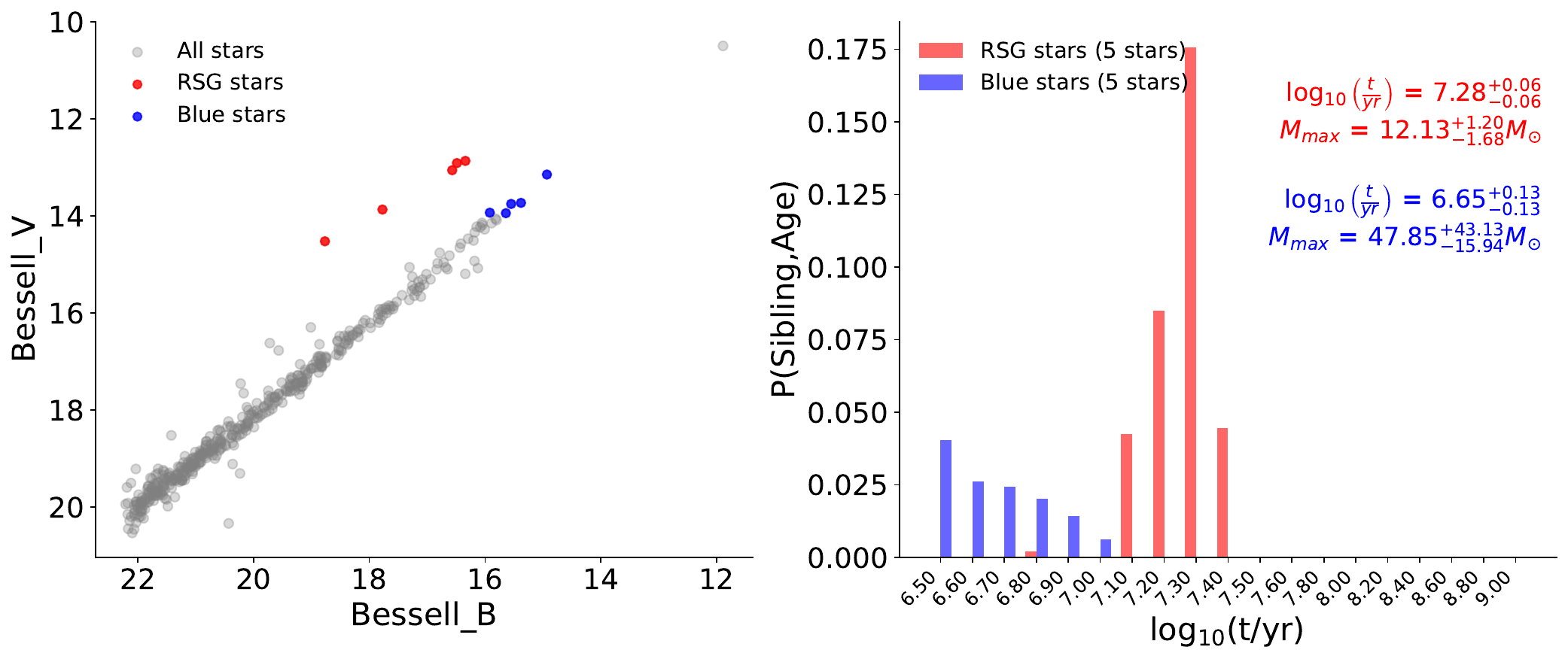}
    \vspace{0.3cm}

    \includegraphics[width=0.83\textwidth]{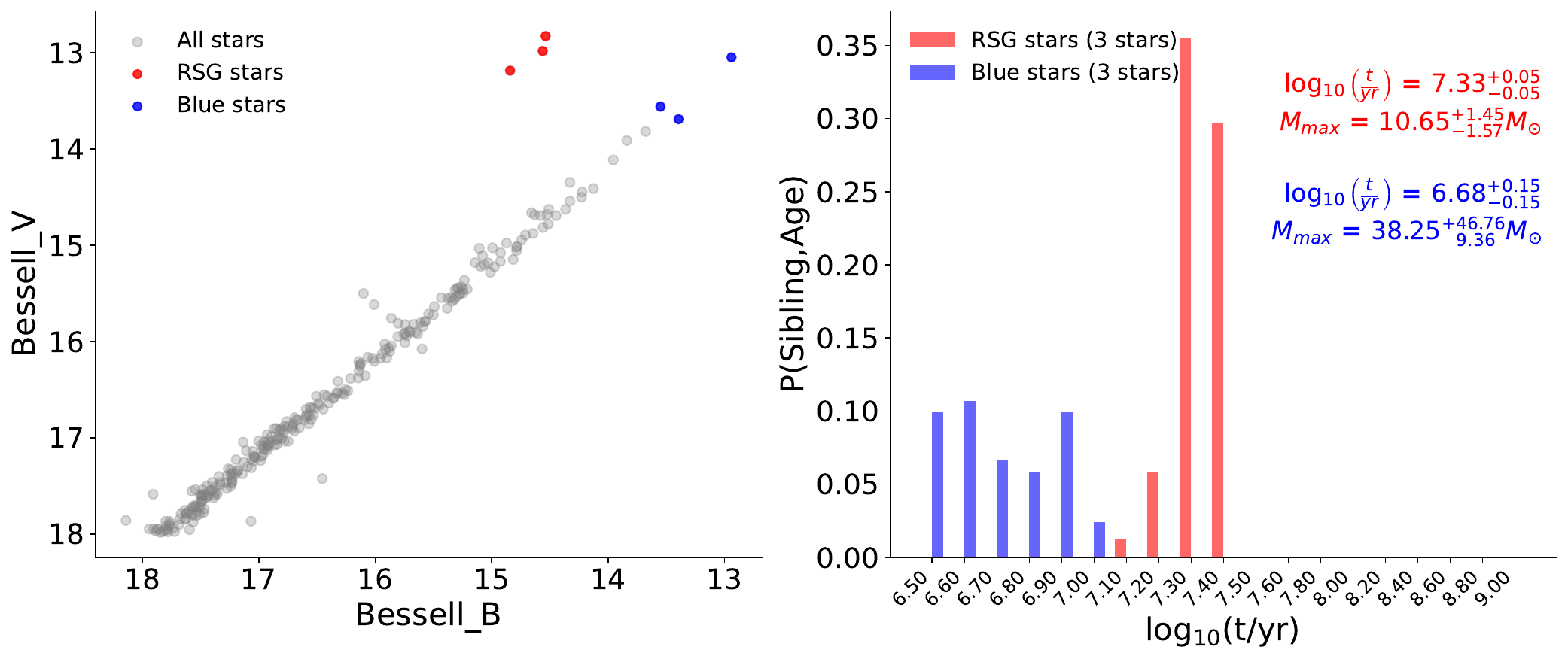}

    \caption{\small Posterior age distributions for selected stellar subsets in NGC 2004 (top), NGC 7419 (middle), and NGC 2100 (bottom), constructed from MCMC draws where at least half of the selected stars are coeval siblings. Each row shows a magnitude–magnitude diagram (left) with the chosen stars highlighted and the corresponding inferred age distributions (right). Red points and bars denote RSGs, while blue indicate bright blue stars. Across all three clusters, the RSGs yield consistent ages of $\log_{10}(t/{\rm yr})\approx7.3$ (18–25 Myr) with narrow, statistically significant posteriors, whereas the bright blue stars favor younger ages of $\log_{10}(t/{\rm yr})\approx6.6$–6.8 (4–6 Myr) but with broad, low-weight distributions. This contrast suggests the RSGs trace a coeval cluster population, while the younger bright blue stars likely reflect model shortcomings or non-standard evolutionary channels.}
    \label{fig:clusters_siblings}
\end{figure*}

The unique capabilities of Stellar Ages to infer ages for both individual stars and entire populations also enables comparisons between tracer subgroups. Subsets of age-sensitive stars can be analyzed separately to test for consistency and reveal systematic differences. Figure \ref{fig:clusters_siblings} illustrates this approach by showing the posterior age distributions obtained when RSGs and bright blue stars are analyzed independently in each cluster. Each panel highlights the chosen stars on a magnitude–magnitude diagram and compares their inferred age distributions. The figure makes clear that RSGs consistently favor ages of $\sim$18–25 Myr, whereas the bright blue stars point to much younger ages of only $\sim$5–6 Myr, immediately signaling a significant discrepancy between the two tracers.

To derive these tracer-based age estimates, we select RSGs and an equal number of bright blue stars as stellar-age tracers. We then compute posterior distributions for each subgroup by combining their individual MCMC draws under the assumption of shared population parameters. This sibling constraint improves age precision by leveraging multiple stars and filtering out draws inconsistent with coevality.

We define $\widehat{P}_{\rm sib}(t)$ as the frequency of all MCMC draws in which at least half of the selected stars occupy age bin $t$. For each draw we count the number of selected stars in each bin and assign a value of 1 to bin $t$ if the count reaches or exceeds $0.5 \times N_{\rm selected}$, and 0 otherwise. Averaging this indicator over the full chain yields $\widehat{P}_{\rm sib}(t)$, which records how often the subset of stars jointly satisfies the sibling condition at bin $t$. The percentile ages are obtained by forming the cumulative distribution over $t$ from these frequencies and reading the 25th, 50th, and 75th percentiles. The formal definition is given in Equation~\ref{eq:Psib}:

\begin{equation}
\label{eq:Psib}
\begin{aligned}
\widehat{P}_{\rm sib}(t) &=
\frac{1}{N_{\rm total}}
\sum_{s=1}^{N_{\rm total}}
\mathbf{1}\!\left\{
n_s(t) \;\ge\; 0.5\,N_{\rm selected}
\right\}, \\[6pt]
n_s(t) &\equiv 
\sum_{i=1}^{N_{\rm selected}} 
\mathbf{1}\{\,t_{i,s} = t\,\}.
\end{aligned}
\end{equation}

This approach offers several advantages over individual-star analysis: it reduces the influence of single-star uncertainties, leverages the statistical power of multiple tracers, and filters out solutions inconsistent with the sibling assumption, testing coevality within each subset.

\subsection{Literature Spectroscopic Constraints on the Brightest Stars}
\label{subsec:spectra}

Figure~\ref{fig:clusters_siblings} highlights the brightest RSGs and bright blue stars (BBS) in our sample. To provide empirical context for these tracers, we summarize their photometry and individual stellar parameter inferences in Table~\ref{tab:brightstars_all}. We then compare these photometrically derived properties directly to published spectroscopic observations.

This spectroscopic comparison requires two different approaches based on cluster distance. For the nearby Galactic cluster NGC~7419, we can perform a direct, star-by-star cross-match with historical literature. The Magellanic Cloud clusters NGC~2004 and NGC~2100 present a different challenge. Our data for these systems come from high-resolution Hubble Space Telescope (HST) imaging of the crowded, central cores. Because these dense regions lack historical, ground-based spectroscopic coverage, we cannot perform star-by-star comparisons. Instead, we evaluate our localized HST inferences against the broader, cluster-wide statistics established by ground-based surveys.

\begin{deluxetable*}{llrrrrrlcccc}
\tablecaption{Photometry and Stellar Ages results for the bright stars highlighted in Figure~\ref{fig:clusters_siblings}. We list apparent and derived absolute magnitudes, the posterior summary parameters from Stellar Ages, and a model-matched spectral classification for each star (Section~\ref{subsec:stat sum}). Specifically, SpT is assigned by matching the posterior-median $\log_{10}(T_{\rm eff}/\mathrm{K})$ and $\log_{10}(L/L_\odot)$ to the \citet{deJager1987} calibration grid across luminosity classes V, IV, III, II, Ib, Iab, Ia, and Ia+}. Tracer labels are RSGs and BBS (bright blue stars). Reported uncertainties are the 25th--75th percentile offsets from the Stellar Ages posterior.\label{tab:brightstars_all}
\tabletypesize{\scriptsize}
\setlength{\tabcolsep}{3pt}
\tablehead{
\colhead{Cluster} &
\colhead{Tracer} &
\colhead{$B$} &
\colhead{$V$} &
\colhead{$B\!-\!V$} &
\colhead{$M_B$} &
\colhead{$M_V$} &
\colhead{SpT$^\mathrm{a}$} &
\colhead{$\log_{10}(T_{\rm eff}/\mathrm{K})$} &
\colhead{$\log_{10}(L/L_\odot)$} &
\colhead{$\log_{10}(t/\mathrm{yr})$} &
\colhead{$M/M_\odot$}
}
\startdata
\multicolumn{12}{c}{\textbf{NGC 2004}, Extinction correction by\citet{Niederhofer2015}, uniform treatment, cluster mean $\bar{A}_{V}\approx0.71$} \\
\tableline
NGC 2004 & RSG & 13.95 & 12.82 & 1.13 & -4.54 & -5.67 & K0 Iab-Ib & $3.673_{-0.007}^{+0.002}$ & $4.300_{-0.008}^{+0.020}$ & $7.347_{-0.027}^{+0.028}$ & $10.012_{-0.007}^{+0.390}$ \\
NGC 2004 & RSG & 13.96 & 12.76 & 1.20 & -4.54 & -5.74 & K0 to K1 Iab-Ib & $3.669_{-0.008}^{+0.005}$ & $4.320_{-0.027}^{+0.027}$ & $7.340_{-0.030}^{+0.030}$ & $10.155_{-0.149}^{+0.253}$ \\
NGC 2004 & RSG & 14.02 & 12.63 & 1.39 & -4.47 & -5.86 & K3 Iab-Ib to M10 III & $3.623_{-0.008}^{+0.020}$ & $4.540_{-0.103}^{+0.055}$ & $7.241_{-0.068}^{+0.066}$ & $11.929_{-1.510}^{+1.415}$ \\
NGC 2004 & RSG & 13.91 & 12.74 & 1.17 & -4.59 & -5.75 & K0 to K1 Iab-Ib & $3.673_{-0.009}^{+0.006}$ & $4.301_{-0.011}^{+0.045}$ & $7.338_{-0.032}^{+0.032}$ & $10.155_{-0.139}^{+0.244}$ \\
NGC 2004 & RSG & 13.91 & 12.91 & 1.00 & -4.59 & -5.58 & K0 Iab-Ib & $3.676_{-0.002}^{+0.003}$ & $4.291_{-0.001}^{+0.009}$ & $7.352_{-0.027}^{+0.027}$ & $10.017_{-0.007}^{+0.361}$ \\
NGC 2004 & BBS & 13.35 & 13.55 & -0.20 & -5.15 & -4.95 & O8 Ia+ to B1 V & $4.489_{-0.085}^{+0.070}$ & $5.120_{-0.157}^{+0.106}$ & $6.695_{-0.158}^{+0.163}$ & $27.159_{-5.988}^{+3.674}$ \\
NGC 2004 & BBS & 13.15 & 13.30 & -0.15 & -5.34 & -5.19 & O8 Ia+ to B1 V & $4.482_{-0.085}^{+0.067}$ & $5.201_{-0.139}^{+0.115}$ & $6.676_{-0.132}^{+0.155}$ & $28.371_{-5.046}^{+4.978}$ \\
NGC 2004 & BBS & 13.45 & 13.51 & -0.06 & -5.05 & -4.98 & O8 Ia+ to B1 V & $4.481_{-0.087}^{+0.068}$ & $5.062_{-0.165}^{+0.139}$ & $6.727_{-0.153}^{+0.169}$ & $24.783_{-8.270}^{+5.445}$ \\
NGC 2004 & BBS & 13.14 & 13.17 & -0.02 & -5.35 & -5.33 & O8 Ia+ to B1 V & $4.469_{-0.090}^{+0.069}$ & $5.089_{-0.334}^{+0.236}$ & $6.807_{-0.226}^{+0.715}$ & $23.975_{-15.869}^{+7.551}$ \\
NGC 2004 & BBS & 12.95 & 13.03 & -0.07 & -5.54 & -5.47 & O7 Ia+ to B1 V & $4.486_{-0.091}^{+0.068}$ & $5.250_{-0.204}^{+0.157}$ & $6.713_{-0.177}^{+0.153}$ & $28.862_{-6.784}^{+7.098}$ \\
\tableline
\multicolumn{12}{c}{\textbf{NGC 7419}, Uncorrected extinction, inferred distribution median $\tilde{A}_{V}\approx6.2$} \\
\tableline
NGC 7419 & RSG & 17.77 & 13.87 & 3.91 & -2.70 & -4.61 & K6 Ia-Iab to M10 III & $3.569_{-0.018}^{+0.040}$ & $4.584_{-0.062}^{+0.099}$ & $7.215_{-0.096}^{+0.063}$ & $11.935_{-0.250}^{+1.583}$ \\
NGC 7419 & RSG & 16.34 & 12.86 & 3.48 & -4.13 & -5.62 & K6 Iab to M10 III & $3.578_{-0.021}^{+0.027}$ & $4.477_{-0.068}^{+0.098}$ & $7.256_{-0.072}^{+0.076}$ & $11.481_{-1.084}^{+1.487}$ \\
NGC 7419 & RSG & 16.49 & 12.91 & 3.58 & -3.99 & -5.57 & K6 Iab to M10 III & $3.571_{-0.019}^{+0.033}$ & $4.532_{-0.029}^{+0.148}$ & $7.225_{-0.095}^{+0.058}$ & $11.860_{-0.498}^{+1.611}$ \\
NGC 7419 & RSG & 16.57 & 13.05 & 3.51 & -3.91 & -5.42 & K5 Iab to M10 III-IV & $3.592_{-0.028}^{+0.026}$ & $4.462_{-0.084}^{+0.141}$ & $7.243_{-0.072}^{+0.059}$ & $11.482_{-0.970}^{+1.846}$ \\
NGC 7419 & RSG & 18.77 & 14.52 & 4.25 & -1.71 & -3.96 & K7 Ia-Iab to M9 II-III & $3.567_{-0.021}^{+0.022}$ & $4.712_{-0.034}^{+0.121}$ & $7.157_{-0.081}^{+0.084}$ & $13.710_{-1.770}^{+2.017}$ \\
NGC 7419 & BBS & 15.92 & 13.93 & 1.99 & -4.56 & -4.55 & O6 Ia+ to A8 Ia-Iab & $4.489_{-0.606}^{+0.091}$ & $5.344_{-1.642}^{+0.253}$ & $6.730_{-0.191}^{+0.878}$ & $29.858_{-23.485}^{+14.108}$ \\
NGC 7419 & BBS & 15.55 & 13.75 & 1.80 & -4.92 & -4.73 & O6 Ia+ to B1 V & $4.515_{-0.093}^{+0.070}$ & $5.262_{-0.263}^{+0.174}$ & $6.715_{-0.178}^{+0.183}$ & $28.523_{-7.608}^{+9.632}$ \\
NGC 7419 & BBS & 15.38 & 13.73 & 1.65 & -5.09 & -4.75 & O8 Ia+ to B1 IV-V & $4.509_{-0.085}^{+0.057}$ & $5.060_{-0.231}^{+0.153}$ & $6.744_{-0.156}^{+0.184}$ & $24.596_{-5.017}^{+5.215}$ \\
NGC 7419 & BBS & 14.93 & 13.14 & 1.79 & -5.54 & -5.34 & O6 Ia+ to B1 V & $4.496_{-0.118}^{+0.075}$ & $5.435_{-0.392}^{+0.208}$ & $6.677_{-0.144}^{+0.229}$ & $35.777_{-13.979}^{+9.903}$ \\
NGC 7419 & BBS & 15.64 & 13.94 & 1.70 & -4.83 & -4.54 & O8 Ia+ to B1 IV-V & $4.498_{-0.089}^{+0.064}$ & $5.012_{-0.253}^{+0.156}$ & $6.757_{-0.170}^{+0.229}$ & $23.742_{-6.929}^{+5.645}$ \\
\tableline
\multicolumn{12}{c}{\textbf{NGC 2100}, Extinction correction by\citet{Niederhofer2015}, differential treatment, cluster mean $\bar{A}_{V}\approx0.53$} \\
\tableline
NGC 2100 & RSG & 14.84 & 13.19 & 1.66 & -3.65 & -5.31 & K6 Iab-Ib to M10 III-IV & $3.601_{-0.019}^{+0.008}$ & $4.364_{-0.061}^{+0.115}$ & $7.281_{-0.043}^{+0.055}$ & $10.382_{-0.111}^{+0.922}$ \\
NGC 2100 & RSG & 14.56 & 12.98 & 1.58 & -3.93 & -5.51 & K5 Iab-Ib to M10 III-IV & $3.603_{-0.003}^{+0.016}$ & $4.463_{-0.097}^{+0.023}$ & $7.277_{-0.041}^{+0.054}$ & $11.087_{-0.746}^{+0.377}$ \\
NGC 2100 & RSG & 14.54 & 12.83 & 1.71 & -3.96 & -5.67 & K6 Ia-Iab to M10 III & $3.601_{-0.016}^{+0.003}$ & $4.495_{-0.016}^{+0.113}$ & $7.274_{-0.039}^{+0.051}$ & $11.967_{-0.850}^{+1.093}$ \\
NGC 2100 & BBS & 12.94 & 13.05 & -0.11 & -5.55 & -5.45 & O7 Ia+ to B0 IV-V & $4.514_{-0.072}^{+0.067}$ & $5.268_{-0.153}^{+0.157}$ & $6.671_{-0.128}^{+0.099}$ & $29.267_{-5.858}^{+8.016}$ \\
NGC 2100 & BBS & 13.39 & 13.69 & -0.30 & -5.10 & -4.80 & O8 Ia+ to O9 Ia+-Ia & $4.539_{-0.060}^{+0.035}$ & $5.102_{-0.121}^{+0.091}$ & $6.775_{-0.161}^{+0.125}$ & $26.513_{-4.367}^{+3.566}$ \\
NGC 2100 & BBS & 13.55 & 13.56 & -0.01 & -4.94 & -4.93 & O8 Ia+ to B0 IV-V & $4.513_{-0.066}^{+0.060}$ & $5.038_{-0.150}^{+0.137}$ & $6.733_{-0.162}^{+0.117}$ & $23.923_{-3.713}^{+5.441}$ \\
\enddata
\tablecomments{$^\mathrm{a}$Spectral classifications are inferred from the Stellar Ages posterior ($B$- and $V$-band photometry propagated through MIST isochrones), not from spectroscopy. The nominal SpT corresponds to the best match at the posterior-median $\log_{10}(T_{\rm eff}/\mathrm{K})$ and $\log_{10}(L/L_\odot)$. The accompanying SpT uncertainty reflects the set of best matches obtained when evaluating the lower/upper bounds of those two quantities (all available corner combinations of the asymmetric 25th--75th percentile offsets). This is an error-envelope/sensitivity range, not a formal confidence interval in SpT.}
\end{deluxetable*}

\begin{figure}[ht]
    \centering
    \includegraphics[width=\columnwidth]{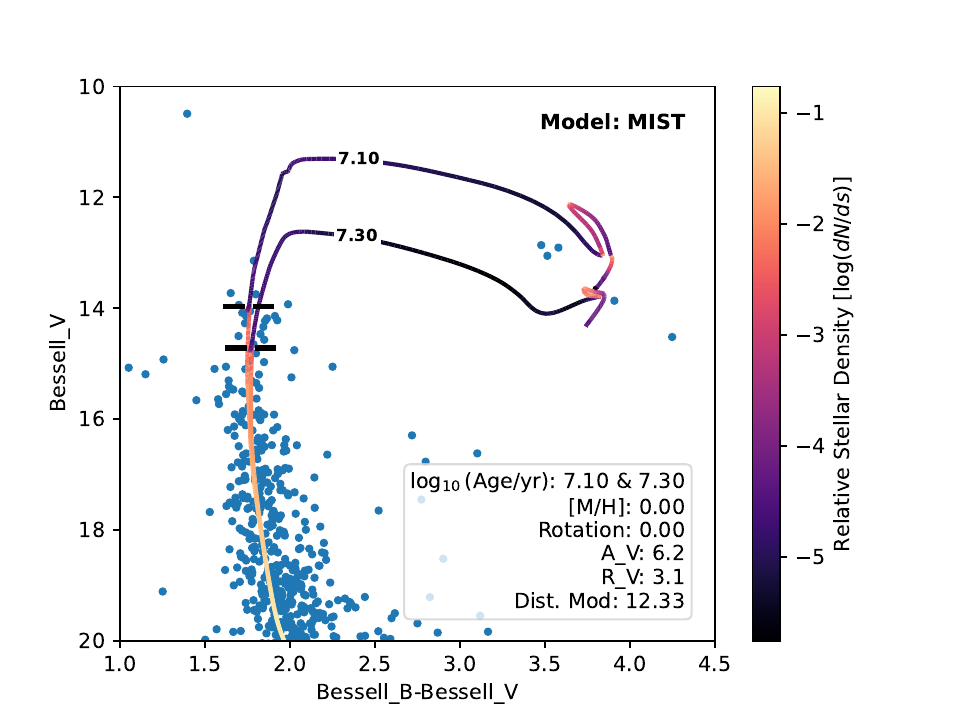}
    \caption{\small CMD for cluster NGC~7419 and MIST isochrones showing two different ages.  The color map shows the relative density of stars assuming a Salpeter IMF.  The horizontal marks indicate the MSTO defined by the drop in stellar density. This demonstrates why the bright blue stars indicate a younger age than the RSGs.  The apparent MSTO is much brighter than is expected for the RSGs.}
    \label{fig:CMDAgeDemo}
\end{figure}

One of the primary results of this study is a stark age discrepancy between the two tracer populations. Specifically, the magnitude distribution of the bright blue stars suggests an age that is much younger than the age inferred from the RSGs. Figure~\ref{fig:CMDAgeDemo} uses the CMD of NGC~7419 to demonstrate why this contradiction occurs. In this diagram, the bright blue stars appear to define the tip of a traditional main sequence turnoff (MSTO). The horizontal dashes mark the position of this practical MSTO, where the stellar density abruptly drops along the isochrone. However, an isochrone with an MSTO bright enough to match these blue stars predicts RSGs that are significantly brighter than any observed in the cluster.

This same discrepancy appears in all three clusters in our sample. The most pressing question is whether the stellar spectra support this apparent physical inconsistency. If they do not, we must identify the underlying modeling or physical biases responsible. In particular, we must determine if the actual observed spectra are consistent with our photometrically inferred effective temperatures and luminosities.

Because the Galactic cluster NGC~7419 has extensive spectroscopic literature for its brightest stars, we begin our analysis here. Table~\ref{tab:ngc7419_spectra} compares our photometrically derived spectral types against historical, spectroscopically derived classifications from \citet{Beauchamp1994_NGC7419}, \citet{Caron_2003}, and \citet{Marco_2013}. Overall, our broad photometry-based spectral ranges are consistent with these direct spectroscopic observations, though the physical drivers for these broad ranges differ at each spectral end. At the blue end, the wide variation stems from observational degeneracy; because the $B$ and $V$ magnitudes of bright blue stars sit on the Rayleigh-Jeans tail, photometry alone cannot easily distinguish precise temperatures. At the population level, our full hierarchical inference breaks this degeneracy by fitting the distribution of stars along the entire blue sequence. Conversely, the wide spectral range at the red end is not caused by photometric degeneracy. Instead, it reflects a physical property of red supergiants: at these low effective temperatures, spectral type has an incredibly shallow dependence on temperature, meaning minor temperature variations encompass a wide range of spectral classifications \citep{Tabernero2018}.

\begin{deluxetable*}{lllll}
\tablecaption{Comparison of Spectral Classifications for Members of NGC 7419 \label{tab:ngc7419_spectra}}
\tablecolumns{5}
\tablewidth{\textwidth}
\tablehead{
\colhead{ID} & \colhead{B94} & \colhead{C03} & \colhead{MN13} & \colhead{Photometric (Ours)}
}
\startdata
\cutinhead{Red Supergiants (RSGs)}
B139 & M3.5 I & \dots & M1 Iab & K8 Ia to M9 II-III \\
B435 & M2 Iab & \dots & M1.5 Iab & K6 Iab to M10 III \\
B696 & M2 Iab & \dots & M1.5 Iab & K7 Ia-Iab to M9 II-III \\
B921 & M2 Iab & \dots & M0 Iab & K7 Iab to M10 III \\
B950 & M7.5 I & \dots & M7.5 I & K8 Ia+-Ia to M8 II-III \\
\midrule
\cutinhead{Bright Blue Stars}
B190 & \dots & B0 III & BN0.5 III & O8 Ia+ to B1 V \\
B389 & B1--2 I:e & Be & B1 IIIe & O8 Ia+ to B1 V \\
B417 & \dots & B1 III--Ve & B1.5 IVe & O8 Ia+ to B1 IV-V \\
B473 & BI & B2.5 III & BC2.5 Ib--II & O8 Ia+ to B0 IV-V \\
B687 & B1--2 I & B2.5 II--III & BC2 II & O8 Ia to B1 V \\
B1129 & \dots & B4 III(e) & B1.5 IV+?(e) & O8 Ia+ to B0 IV-V \\
\enddata
\tablecomments{B94: Beauchamp et al. (1994); C03: Caron et al. (2003); MN13: Marco \& Negueruela (2013). Note that stars are listed in the identical cross-matched sequence as presented in Table~\ref{tab:brightstars_all}.}
\end{deluxetable*}

Despite these broad ranges, the literature and our work remain mostly consistent regarding the cluster's red supergiants. Both approaches classify these stars as early- to mid-M supergiants with stable photometric behavior and a shared average luminosity of $\log_{10}(L/L_\odot) \approx 4.6$. One notable exception in the literature is MY~Cep, an exceptionally late-type, luminous M7.5 supergiant. Because MY~Cep exhibits extreme circumstellar extinction and high variability, we explicitly excluded it from our RSG tracer sample to prevent a single peculiar object from dominating the inferred cluster age signal \citep{Verheyen2012, Chakraborty_2025}. The remaining five RSGs present stable photometry and consistent M-type classifications, indicating that our photometrically derived red supergiant luminosities match historical spectroscopic expectations.

In contrast, the historical classification and luminosity determination for the bright blue stars are far more varied and uncertain. Historically, narrative bias has often driven conclusions for these blue stars, forcing them to fit the older ages implied by the RSGs. For example, the original classifications by \citet{Beauchamp1994_NGC7419} assigned these stars a supergiant (I) luminosity class. Later, \citet{Caron_2003} used equivalent widths to infer surface gravity ($\log g$) and subsequently downgraded these classifications to luminosity classes II and III. However, we urge caution in adopting either interpretation. NGC~7419 hosts a famously high fraction of Be stars, which indicates exceptionally rapid stellar rotation. In equivalent width measurements, rapid rotation creates a direct physical degeneracy with $\log g$. While line-profile shapes can theoretically break this degeneracy, \citet{Caron_2003} did not report the goodness-of-fit for their line shapes. Furthermore, while they convolved their spectral models with rotation, they did not specify the rotation rates tested. At the time, literature models commonly defaulted to a modest fractional rotation of $v/v_{\rm crit} \approx 0.4$, a value known to leave visual line properties largely unmodified. Because actual B-star populations extend to much higher near-breakup rotation rates, historical spectroscopic models that neglect extreme rotation will systematically miscalculate $\log g$.

Because our own photometric models also omit higher values of rotation, we are effectively comparing unrotated models against unrotated models. Unsurprisingly, our photometrically derived types align closely with the \citet{Caron_2003} classifications. More recent work highlights the persistent uncertainty surrounding these stars. \citet{Marco_2013} upgraded the brightest blue star back to a Ib--II classification, while noting that the rest of the blue population spans luminosity classes V to III. Crucially, to force a consistent single-star age between the blue sequence and the RSGs, \citet{Marco_2013} omitted all Be stars from their final analysis. By discarding the brightest stars on the blue sequence, their methodology inadvertently demonstrates that this age inconsistency has been clear for decades. This unique population—characterized by an abundance of RSGs, a lack of classical blue supergiants, and a massive Be-star fraction—strongly suggests that rotation and binary interactions dictate the post-main-sequence evolution of many of the massive stars in NGC~7419. Ultimately, because historical spectroscopic classifications rely on models lacking rapid rotation, their resulting spectral types remain highly uncertain.

The remaining two clusters in our sample do not possess the same extensive star-by-star spectroscopic literature as NGC~7419. While historical studies of NGC~2004 and NGC~2100 have successfully targeted and classified their red supergiant populations, neither cluster has been the subject of a dedicated, individual classification survey for its blue stellar sequence. We therefore rely on broader, cluster-wide statistical results from large-scale initiatives like the VLT-FLAMES survey to provide spectroscopic context.

The Large Magellanic Cloud (LMC) cluster NGC~2004 contains a rich mixture of highly luminous red and blue stars. Several of the cluster's bright red members are spectroscopically confirmed as K-type supergiants. For instance, \citet{Feast1979} obtained spectra for four of these RSGs, classifying them as K0--K1~Ib stars. These direct physical classifications align closely with the photometrically inferred parameters we present in Table~\ref{tab:brightstars_all}. The literature also identifies a highly luminous emission-line composite system within the cluster, interpreted as a VV~Cephei-like interacting binary (R~108) \citep{Henize1956}. Beyond these peculiar binary systems, the blue stellar sequence of NGC~2004 harbors massive stars that independently suggest a very young cluster age. The VLT-FLAMES survey, for example, identified at least one early O-type member designated as NGC~2004-049 \citep{Evans2006}. Published ultraviolet spectra and classification work on the upper main sequence confirm that a dense population of hot O- and B-type stars defines this blue sequence. Several of these stars appear brighter and bluer than traditional single-star main sequence turn-offs, a feature that \citet{Grebel1996} argued is a natural consequence of binary evolution.

In contrast, NGC~2100 exhibits an even more extreme dominance of red supergiants at the bright end of its color-magnitude diagram. Integrated-light and infrared studies indicate that NGC~2100 hosts between 15 and 20 RSGs, though the spatial constraints of our specific HST aperture restrict our active analysis to a smaller subset \citep{Beasor2016_NGC2100RSGs, Patrick2016}. Spectroscopic investigations classify several of these members as early- to mid-M supergiants. Similar to NGC~2004, the literature reports at least one interacting VV~Cephei-like binary system among these bright evolved stars \citep{Robertson1974, Feast1979}. Conversely, the luminous blue stellar content of NGC~2100 is comparatively sparse. Its apparent main sequence turn-off sits near spectral types B1--B2, and its most luminous blue members are restricted to mid-B types. Furthermore, historical ultraviolet studies do not detect extremely luminous OB supergiants extending past this turn-off \citep{Bohm-Vitense1985}. The spectroscopic literature thus paints a consistent picture of NGC~2100: its upper color-magnitude diagram is dominated by a massive cohort of red supergiants, with relatively few hot, post-main-sequence blue stars.

Ultimately, the historical spectroscopic data available for all three clusters support the age discrepancies we quantify in this work. Where detailed spectra exist, they either reflect the same physical tensions seen in our photometric analysis or highlight the severe systematic uncertainties inherent in classifying rapidly rotating Be stars.

\subsection{Statistical Summary of Age Discrepancies}
\label{subsec:stat sum}

Table~\ref{tab:cluster_ages} presents the inferred ages and maximum stellar masses ($M_{\rm max}$) for the RSG and bright blue star populations across the three young clusters. Interestingly, there are some consistent patterns in the age discrepancy.

\renewcommand{\arraystretch}{1.3} 
\begin{deluxetable}{l c c c c c}
\tabletypesize{\scriptsize}
\tablecaption{Cluster age estimates from RSG and bright blue star populations. This table summarizes the results of figure \ref{fig:clusters_siblings}. \label{tab:cluster_ages}}
\tablehead{
\colhead{Cluster} &
\colhead{$N_{\rm RSG}$} &
\colhead{$N_{\rm B}$} &
\colhead{$\log_{10}(t/{\rm yr})$} &
\colhead{$t$ (Myr)} &
\colhead{$M_{\rm max}$ [$M_\odot$]} \\
}
\startdata
NGC~2004 & 5 & 0 & $7.40^{+0.05}_{-0.05}$ & $25.1^{+3.0}_{-2.7}$ & $8.8^{+1.4}_{-0.1}$ \\
         & 0 & 5 & $6.70^{+0.16}_{-0.16}$ & $5.0^{+2.0}_{-1.4}$ & $34.1^{+47.1}_{-7.1}$ \\
NGC~7419 & 5 & 0 & $7.28^{+0.06}_{-0.06}$ & $19.1^{+2.8}_{-2.5}$ & $12.13^{+1.20}_{-1.68}$ \\
         & 0 & 5 & $6.65^{+0.13}_{-0.13}$ & $4.5^{+1.6}_{-1.2}$ & $47.85^{+43.13}_{-15.94}$ \\
NGC~2100 & 3 & 0 & $7.33^{+0.05}_{-0.05}$ & $21.4^{+2.6}_{-2.3}$ & $10.65^{+1.45}_{-1.57}$ \\
         & 0 & 3 & $6.68^{+0.15}_{-0.15}$ & $4.8^{+1.9}_{-1.4}$ & $38.25^{+46.76}_{-9.36}$ \\
\enddata
\tablecomments{
$N_{\rm RSG}$ and $N_{\rm B}$ give the number of RSGs and bright blue stars, respectively, used in each age estimate. The uncertainties reported correspond to the 25\% and 75\% quartiles of the posterior $\log_{10}$(t/yr) and M$_{max}$.
Linear ages in Myr are obtained by exponentiating the $\log$ age quantiles, yielding asymmetric uncertainties in linear space.
$M_{\rm max}$ is in units of M$_{\odot}$ and represents the most massive star implied by age.
}
\end{deluxetable}
\renewcommand{\arraystretch}{1.0}

While tempting, directly comparing these two populations results would not be entirely self-consistent. The RSG-derived ages assume that the RSGs are siblings. The bright-blue-star-derived ages assume that the bright blue stars are siblings, and this assumption leads to selecting \textit{different} samples from the MCMC draws. Although each subset yields self-consistent inferences when considered in isolation, comparing the two populations based on different underlying draws would introduce a bias. To address this, we adopt an alternative test designed to enforce self-consistency throughout the analysis. Specifically, we require the RSGs to satisfy the sibling condition and then retain the corresponding MCMC samples for the bright blue stars simultaneously. For each such draw, we compute the differences in ages, and likewise the ratios of maximum stellar masses, ensuring that the blue star inferences are conditioned on the same underlying population properties as the RSGs. All reported discrepancies below are therefore based on these consistent samples. In practice, the differences between the direct comparison and consistent-draw comparisons are modest, but the latter provides a more reliable basis for quantifying discrepancies. 

The discrepancy in RSG-derived and bright-blue-star-derived ages is similar among the three clusters:

\begin{equation}
\Delta \log_{10}(t/{\rm yr}) =
\begin{cases}
0.60^{+0.10}_{-0.10} & \text{NGC~2004}, \\
0.50^{+0.10}_{-0.18} & \text{NGC~7419}, \\
0.60^{+0.10}_{-0.10} & \text{NGC~2100}. \\
\end{cases}
\end{equation}
where $\Delta \log_{10}(t/{\rm yr})$ is the difference in ages between the estimates: $\log_{10}(t_{RSG}/{\rm yr}) - \log_{10}(t_{B}/{\rm yr})$, and the reported uncertainties denote the 25th and 75th percentiles relative to the median. 
The median of these differences is:

\begin{equation}
\Delta \log_{10}(t/{\rm yr})_{\text{\tiny med}} = 0.60 \pm 0.06~{\rm dex} \, ,
\end{equation}

where the uncertainty is based on the range of median log age differences across the clusters.
This log age difference corresponds to a multiplicative age factor of
\begin{equation}
t_{B} \approx \frac{t_{RSG}}{4.0}
\end{equation}
with a range of $[3.2, 4.0]$ across the cluster medians.

Expressed in linear age differences, the resulting median values are:

\begin{equation}
\Delta t_{med} =
\begin{cases}
18.81^{+1.30}_{-1.63}~{\rm Myr} & \text{NGC~2004}, \\
12.01^{+2.93}_{-2.47}~{\rm Myr} & \text{NGC~7419}, \\
15.12^{+3.69}_{-1.48}~{\rm Myr} & \text{NGC~2100}. \\
\end{cases}
\end{equation}

The median across the clusters is $\Delta t_{\text{\tiny med}} = 15.12 \pm 3.93~{\rm Myr}$, again using the range-based uncertainty estimate.

To further quantify the discrepancy between the stellar subpopulations, we examined the maximum stellar masses implied by the RSG and blue star ages. Using the consistent-draw method described above, we find that the mass ratios are high.

\begin{equation}
\frac{M_{\rm max,BBS}}{M_{\rm max,RSG}} =
\begin{cases}
3.6^{+0.32}_{-0.83} & \text{NGC~2004}, \\
2.59^{+0.99}_{-0.62} & \text{NGC~7419}, \\
2.92^{+2.55}_{-0.25} & \text{NGC~2100}. \\
\end{cases}
\end{equation}
The median ratio across clusters is $\left(M_{\rm max,BBS}/M_{\rm max,RSG}\right)_{\text{\tiny med}} = 2.92 \pm 0.58$, with a range-based uncertainty.

The direct-comparison ratios suggest similar values of roughly 3–4 times more massive blue stars. Across all three clusters, blue-star ages imply maximum stellar masses a factor of 2.6–3.6 larger than those inferred from RSGs. This reinforces the conclusion that the two populations yield systematically discrepant mass inferences.

In summary, the discrepancy is systematic across all three clusters: blue stars appear $\sim 4$ times younger, corresponding to a difference of $\sim 15$ Myr younger. The corresponding maximum mass is $\sim 3$ times larger than the RSG-based values. These findings challenge single-star interpretations of coeval populations and indicate that additional physics, such as rapid rotation, binary interaction, or stellar rejuvenation, may be required to explain the observed differences.

\section{Interpretation}
\label{sec:Interpretation}

We first compare our findings with the age discrepancies identified by \citet{Beasor2019} in Section~\ref{subsec:Beasor}. We then examine the implications for stellar evolution models, cluster age dating, population normalization, and UV-based star-formation estimates in Section~\ref{subsec:implications}. Finally, Section~\ref{subsec:Binarity} considers binary interaction, stellar rejuvenation, and rapid rotation as possible explanations for the anomalously young ages inferred from the brightest blue stars.

\subsection{Revisiting Beasor}
\label{subsec:Beasor}

\citet{Beasor2019} presented one of the first systematic demonstrations that different age tracers are leading to discrepant ages in young stellar clusters. By comparing three independent methods, the brightest main-sequence turnoff (TO) star, the TO luminosity function, and the lowest luminosity RSG, they found systematic discrepancies of up to a factor of two in inferred cluster ages. In particular, the TO-based methods consistently yielded younger ages (typically 7–10 Myr), while the lowest-luminosity RSGs implied significantly older ages (20–24 Myr in the Magellanic Cloud clusters). \citet{Beasor2019} argued that the most luminous stars above the TO are unlikely to represent a different coeval population and are more plausibly explained by a combination of binary interaction, rapid rotation, and other effects not captured by single-star models.  Our results reinforce this qualitative age discrepancy as well as quantify it.

In addition to illustrating the discrepancy in ages, we discuss how this impacts inferred maximum masses as well. Although \citet{Beasor2019} framed the discrepancy primarily in terms of age diagnostics, the same age offsets naturally translate into differences in inferred maximum initial masses. Our results show that the younger blue-star ages imply maximum stellar initial masses roughly three times higher than those inferred from the RSGs. This mass tension reinforces the conclusion that the brightest blue stars cannot be explained by standard single-star evolution at the cluster age and likely trace evolutionary pathways or physical processes which are absent from current isochrone models \citep{Schneider2014}.

Together, our results confirm, sharpen and point to possible explanations for the conclusions of \citet{Beasor2019}. The systematic offset between TO-based and RSG-based ages is real and persistent across different clusters. By quantifying these discrepancies in a Bayesian framework, we strengthen the case that current models of very young clusters contain systematic discrepancies, with the brightest blue stars highlighting where single-star assumptions tend to break down.

\subsection{Implications for Stellar Evolution Models}
\label{subsec:implications}

Our analysis confirms a systematic and significant discrepancy between ages inferred from RSG populations and those derived from the brightest blue stars. Across all three clusters, the typical difference is $\sim0.6$ dex in log age, which corresponds to a factor of three to four in linear time and predicts maximum stellar masses that are systematically lower by a similar factor. These results suggest that current stellar evolution models, even those incorporating moderate rotation (e.g. MIST \citet{MIST2016_0, MIST2016_1}), do not fully capture the CMDs of very young clusters.  This implies missing physics such as very rapid rotation and/or binarity, among other processes.

Although the broadened blue sequences in these clusters resemble the extended main-sequence turnoffs (eMSTOs) observed in intermediate-age systems \citep{Mackey2007, Niederhofer2016}, their detailed morphology differs both qualitatively and quantitatively. In intermediate-age clusters, rapid rotation modifies stellar brightness and temperature, smearing the MSTO while leaving the feature recognizable \citep{Bastian2009}. In massive, very young clusters, however, several mechanisms conspire to obscure the turnoff \citep{Li2019}. Beyond rotation and inclination, which can shift stars by up to order a magnitude in color and brightness \citep{Wang2023}, binary evolution can also play a critical role \citep{HSana2012, deMink2014}. Mass transfer and mergers generate rejuvenated stars that populate the upper main sequence, further complicating the turnoff region \citep{Schneider2016}. Furthermore, post-main-sequence evolved stars, such as blue supergiants and some Wolf-Rayet progenitors, can evolve back to the blue sequence \citep{Crowther2007, Ekstrom2012_Geneva, Shara_2013}. Collectively, these effects significantly brighten the apparent MSTO, mimicking a younger age and rendering single-star isochrones insufficient for accurate interpretation \citep{Gossage_2019}.

Compared to the broader main sequence, the RSG-based ages are the only ones that achieve relative self-consistency with the observed main-sequence age distribution. In contrast, the brightest blue stars favor much younger ages that lack the expected population of lower-mass main sequence stars under a standard IMF. This is consistent with a population that likely includes mass gainers, mergers or rapid rotators \citep{Schneider2014, Beasor2019}

A modest intrinsic age spread may also contribute to the smearing of the blue sequence. Massive stars evolve on timescales of millions of years \citep{Ekstrom2012_Geneva, MIST2016_1}. Therefore an age dispersion of one to two million years can produce visible broadening in a CMD, even though it falls far short of explaining the observed ten to twenty million year RSG–blue star age discrepancy. Age spreads of order a few million years are expected in realistic star-forming environments and likely add minor structure to the blue sequence \citep{Krumholz_2007, DaRio_2010}, but they cannot fully account for the systematic offset in age tracers that we observe.

A distinct result emerges from number counts. For an observed number of RSGs in each cluster, single-star models predict more main-sequence stars than we observe, with over predictions by factors of a few. This does not necessarily imply that the RSG ages are incorrect. It points instead to a mismatch in the modeled conversion from RSG incidence to main-sequence counts. Possible causes include RSG lifetimes that are too short in the models, mass-loss prescriptions that are too strong \citep{Declin2024, Gormaz-Matamala2024, Zapartas2025}, or pre-RSG binary channels that increase the number of RSGs relative to single-star expectations \citep{Meynet2015,Beasor2016_NGC2100RSGs}. Environmental effects in young, dense clusters and uncertainties in rotational mixing may also contribute. Photometric incompleteness of the main sequence is another possible contributor, although completeness tests and the consistency of the result across clusters make this unlikely to dominate. Finally, small-number stochasticity is important because adding even one or two RSGs can significantly shift the inferred mapping and amplify the apparent tension.

These findings illustrate that the apparent inconsistencies are not random scatter but physically motivated discrepancies in how massive-star populations are modeled. RSGs remain valuable age tracers because they yield ages that are consistent with the main sequence in our data and are less sensitive to certain binary pathways than other evolved phases. This conclusion is reinforced by the modeling work of \citet{Eldridge2020}, who used \texttt{BPASS} \citep{Eldridge2017} to show that RSG-based ages remain essentially unchanged even when interacting binaries are included, and confirmed this result in NGC~2004, NGC~7419 and NGC~2100. However, we found that specifically converting RSG counts into main-sequence normalizations with current single-star models can bias the inferred main-sequence population. Population synthesis, IMF inference, and feedback calibrations should cross-check ages and normalizations using multiple tracers and, where possible, use models that include binary evolution and rotation \citep{Conroy2013}. For clusters like those studied here, we recommend anchoring ages to the RSGs while treating the brightest blue stars as a composite population influenced by binary interaction, rapid rotation, and possible brief age spreads or WR evolution. These RSG-based ages are therefore the most reliable benchmark available at present, while our quantified discrepancies underscore the importance of improving models so that the ages of bright blue stars can eventually be interpreted on the same footing.

A related implication concerns UV-based estimates of recent star formation. Because the bright blue stars that dominate the UV flux in young populations yield ages a factor of three to four younger than the RSGs, our results suggest that UV-derived star-formation rates may be biased high if rejuvenated or rapidly rotating stars are not accounted for. At the same time, \citet{Choi_2025} showed that including UV data alongside optical CMDs improves the precision of recent star-formation histories by reducing misclassification of evolved stars and tightening uncertainties by about 5-20\%. These results suggest that while broader wavelength coverage can reduce observational uncertainty, the larger systematic offsets we identify point to the need for stellar population models that incorporate binary and rapid rotation channels.

\subsection{Possible Resolutions}
\label{subsec:Binarity}

The systematic age offset could represent a combination of effects, including binary interactions, stellar rejuvenation, and rapid rotation, any of which can produce apparently younger bright blue stars.
Binary evolution can produce stars that appear younger and more massive than single-star models predict through mass transfer, mergers, or binary induced rotation \citep{Eldridge2009, deMink2014}. Tidal interactions in close binaries can spin up a star and drive strong rotational mixing. This mixing grows the core and extends the main sequence lifetime, which makes the star brighter and gives it the appearance of a younger object at fixed initial mass. These processes naturally create bright blue stars that are anomalously young compared to the surrounding population and are consistent with the ages and maximum masses inferred in our analysis.

Rejuvenated stars formed through mass accretion or mergers, may represent a significant fraction of the bright blue star population \citep{Sandage1953, Ferraro2009}. Their presence can bias age estimates toward younger values. In the clusters discussed in this paper, the substantial difference between blue star and RSG ages implies that binary evolution and/or rapid rotation contribute significantly to the observed population. This interpretation aligns with previous studies indicating that massive star clusters can host a high fraction of binary systems, particularly in dense environments where dynamical interactions are frequent \citep{HSana2012, Moe2017}.

In addition to binary channels, rapid stellar rotation can significantly alter the observed luminosities and temperatures of stars, which in turn affects inferred ages. Rotation modifies stellar structure and evolution by enhancing internal mixing, extending main-sequence lifetimes, and modifying luminosities and temperatures at a given mass \citep{Maeder2000, Heger_2000}. Inclination further complicates the picture. Gravitational darkening causes rapidly rotating stars to appear brighter and hotter when viewed pole on and fainter and cooler when viewed equator on \citep{Zeipel1924}. Population synthesis models often adopt moderate rotation rates (typically $\omega / \omega_{\rm crit} \sim 0.4$) \citep{Ekstrom2012_Geneva}, yet observations of Be stars show that many massive stars can rotate much faster \citep{porter2003}. In fact, some Be stars rotate at or near the critical rate, although gravity darkening causes spectra to underestimate their true rotational velocities because the cooler equatorial regions contribute little to the observed flux \citep{Townsend2004}.
Such extreme rotation can make stars appear significantly younger and more luminous than non-rotating counterparts of the same mass. Recent work has found that a strikingly high fraction of stars in several Magellanic Cloud clusters are fast rotators, with initial rotation rates above $\omega / \omega_{\rm crit} > 0.7$ for the majority of the population \citet{ettorre2025}. As argued in \citet{Murphy_2025}, incorporating a broader distribution of rapid rotation rates into stellar models may help reconcile tensions between observed populations and standard single-star predictions. In the context of the clusters studied here, rapid rotation could therefore provide an additional mechanism, alongside binarity, for producing anomalously young-appearing bright blue stars.

Accounting for these effects requires both observational and modeling considerations. Spectroscopic surveys can constrain multiplicity fractions and orbital properties, while population synthesis models that include binary channels can predict the frequency and properties of rejuvenated stars \citep{Sana2013, Eldridge2017}. Incorporating these effects into age-dating analyses may reconcile the apparent tension between RSG and blue star ages, providing a more complete understanding of cluster stellar populations. Ultimately, the systematic offsets observed here point to an emerging inconsistency in how current models represent the evolution of very young clusters. A combination of incomplete single-star physics, binary-induced rejuvenation, and rotation likely contribute, underscoring the need for models that can sufficiently treat these effects.

\section{Conclusions}
\label{sec:conclusions}

Using the novel statistical algorithm Stellar Ages, our analysis of the nearby clusters NGC~2004, NGC~7419 and NGC~2100, reveals a systematic and quantifiable age discrepancy when considering RSGs and bright-blue stars as age tracers. The brightest blue stars consistently indicate ages that are too young to be reconciled with the bulk of the MS. In contrast, the RSG and lower-mass main-sequence populations yield mutually consistent ages, indicating that RSGs are currently the most reliable tracers of the clusters' underlying age. However, even when adopting RSG-based ages, models still overpredict the number of main-sequence stars, revealing an additional population-level mismatch.

Quantitatively, the difference in inferred ages between RSGs and bright blue stars is $\Delta \log_{10}(t/{\rm yr})_{\text{\tiny med}} = 0.60 \pm 0.0
6$ dex, corresponding to a multiplicative age factor of $t_B \approx t_{\rm RSG}/4.0$. This consistent offset challenges the assumption that a coeval population described by single-star tracks can explain these clusters. Multiple effects, such as binary interaction, mergers, rapid rotation, and possibly modest age spreads likely contribute to the apparent youth of the brightest blue stars, underscoring the incompleteness of current stellar models in this regime.

These results highlight the difficulty of age-dating very young clusters. Although RSGs provide the most internally consistent ages, a complete understanding requires combining multiple tracers, expanding the treatment of rotation and binary evolution in population models, and testing those predictions against CMD statistics. The systematic differences we identify reveal that current models do not fully capture the physics that shape very young stellar populations. Resolving these gaps is necessary to refine massive star evolution and to reconstruct cluster formation histories. These issues also influence the broader interpretation of massive stars. For instance, core-collapse supernova progenitor inferences rely on models of their observed properties, yet our results show that the most luminous stars in young clusters often deviate from the assumptions built into standard single-star tracks. This bias affects the initial masses assigned to these stars and introduces uncertainty into progenitor classifications.

\vspace{0.5cm}

\section*{Acknowledgments}
JWM was supported by the Los Alamos National Laboratory (LANL) through its Center for Space and Earth Science (CSES). CSES is funded by LANL’s Laboratory Directed Research and Development (LDRD) program under project number 20210528CR.  Support for M.A. was provided by the VITA-Origins Fellowship, including funding from the Virginia Institute of Theoretical Astronomy (VITA), supported by the College and Graduate School of Arts and Sciences at the University of Virginia. ERB is supported by a Royal Society Dorothy Hodgkin Fellowship (grant no. DHF-R1-241114). This work was supported by NASA through grant HST-GO-16778.017-A from the Space Telescope Science Institute, which is operated by the Association of Universities for Research in Astronomy, Inc., under NASA contract NAS 5-26555.

\facilities{HST (WFPC2)}

\bibliography{joe}
\bibliographystyle{aasjournalv7}

\end{document}